\documentclass[iop]{emulateapj}
\shorttitle{Nuclear Metallicity of Interacting Galaxies}
\shortauthors{Torrey et al.}
\usepackage{mathtools}
\usepackage{graphicx}
\usepackage{epstopdf}
\usepackage{natbib}
\usepackage{verbatim}

\begin{document}

\def \ltsim {\lesssim}

\title{The Metallicity Evolution of Interacting Galaxies}

\author{Paul Torrey\altaffilmark{$\star$}\altaffilmark{1} }
\author{T. J. Cox \altaffilmark{2}}
\author{Lisa Kewley \altaffilmark{3} }
\author{Lars Hernquist \altaffilmark{1} }

\altaffiltext{1}{Harvard Smithsonian Center for Astrophysics, 60 Garden St., Cambridge, MA 02138}
\altaffiltext{2}{Observatories of the Carnegie Institute of Washington, 813 Santa Barbara St., Pasadena, CA, 91101}
\altaffiltext{3}{Institute for Astronomy, University of Hawaii, Honolulu HI, 96822}
\altaffiltext{$\star$}{Email: ptorrey@cfa.harvard.edu}

\begin{abstract}

Nuclear inflows of metal--poor interstellar gas triggered by galaxy
interactions can account for the systematically lower central oxygen
abundances observed in local interacting galaxies.  Here, we
investigate the metallicity evolution of a large set of simulations of
colliding galaxies.  Our models include cooling, star formation,
feedback, and a new stochastic method for tracking the mass recycled
back to the interstellar medium from stellar winds and supernovae.  We
study the influence of merger--induced inflows, enrichment, gas
consumption, and galactic winds in determining the nuclear
metallicity.  The central metallicity is primarily a competition
between the inflow of low--metallicity gas and enrichment from star
formation.  An average depression in the nuclear metallicity of
$\sim0.07$ is found for gas--poor disk--disk interactions.  Gas--rich
disk--disk interactions, on the other hand, typically have an
enhancement in the central metallicity that is positively correlated
with the gas content.  The simulations fare reasonably well when
compared to the observed mass--metallicity and separation--metallicity
relationships, but further study is warranted.

\end{abstract}

%%%%%%%%%%%%%%%%%%%%%%%%%%%%%%%%%%%%%%%%%%%
%%%%%%%%%%%%%%%%%%%%%%%%%%%%%%%%%%%%%%%%%%%
%%%%%%%%%%%%%%%%%%%%%%%%%%%%%%%%%%%%%%%%%%%

\section{Introduction}

The nuclear metallicities of star-forming galaxies are
characterized by a mass-metallicity relation
\citep[hereafter, MZ][]{Lequeux1979, Rubin1984,Tremonti2004}.  Contributing to the
scatter in the MZ relation are interacting galaxies, which are
consistently lower in central metallicity than non--interacting
systems of equivalent mass, as first noted by \citet{KGB06}, and later
confirmed for ultraluminous infrared galaxies (ULIRGs)
\citep{Rupke2008}, close pairs in SDSS 
\citep{SloanClosePairs,MichelDansac2008,Peeples2009,SolA2010}, and local, low
mass systems \citep{EktaChengalur2010}.  Observations such as these
are most naturally explained as the result of vigorous merger--induced
inflows of gas \citep{BH91,BH96} which rearrange the initial
metallicity gradient \citep{Shields1990,BelleyRoy1992,Z94,MA04} and
``dilute'' the nuclear metallicity \citep{KGB06}.  In this sense, the
same gas that drives nuclear starbursts \citep{MH94a,MH96,Iono04}
and triggers central AGN activity and
black hole growth \citep{DSH05,SDMHModel,HH06,
Hopkins07}, also results in suppressed nuclear
metallicity.

Simulations of merger-driven inflows of gas naturally predict a
flattening of the initial metallicity gradient~\citep{Perez2006,Rupke2010,PerezScoop},
which has now been observed in a number of colliding systems
\citep{KewleyGradients2010,RupkeGradients2010}.  However, more than
hydrodynamics are at play in determining the nuclear metallicity
evolution.  To further our understanding, we should consider ongoing
star formation with associated chemical enrichment, feedback from star
formation and AGN activity, and the interchange of material between
the stellar and gaseous phases as stars are born and later return
material to the interstellar medium.

The four main processes responsible for the evolution of the nuclear
metallicity are gas inflows, chemical enrichment from star formation,
gas consumption, and galactic outflows.  These effects compete with
one another to influence the nuclear metallicity, making it difficult
to determine their relative importance a priori.  Numerical
simulations have only recently been used to quantify the detailed
impact of these effects on metallicity gradients and nuclear
metallicities of interacting pairs.  \citet{Rupke2010} simulated
colliding galaxies without chemical enrichment to explore dynamically
induced changes in metallicity gradients, finding that a drastic
flattening can occur, accompanied by a drop in the nuclear
metallicity.  \citet{DilutionPeak2010} and~\citet{PerezScoop} performed 
simulations with star formation and chemical enrichment and found
similar results to \citet{Rupke2010}, but noted that the dip in the
nuclear metallicity values can be counteracted by chemical enrichment
from star formation.  These simulations have refined our knowledge of
the depressed MZ relation.

In this paper, we explore the evolution of the nuclear metallicity
during mergers using numerical simulations which include cooling, star
formation, stellar feedback, and black hole growth and AGN feedback.
Our approach allows us to systematically investigate the importance of
gas inflows, chemical enrichment from star formation, gas consumption
as a result of star formation, and galactic outflows.  In order to
unambiguously determine the role of metal enrichment, we have
developed a stochastic method to recycle stellar particles back to the
interstellar medium without requiring inter-particle mass mixing.  The
stochastic gas recycling method, which is designed by analogy to the
widely used stochastic star formation method, has attributes that are
distinct from kernel weighted mass return, making it particularly
useful for our study.

We consider a range of situations with systematically varied
parameters to enhance our understanding of metallicity evolution as a
natural extension of the merger process.  We find that, for systems
modeled after those in the local Universe, the main driver behind
changes to the nuclear metallicity is the flow of low metallicity gas
to the nuclear regions, or metallicity dilution.  However, we identify
two previously under-appreciated effects that influence the strength
of this dilution: locking of gas and metals in the stellar phase, and
gas and metal removal via stellar-driven winds.  We also find that,
for systems modeled after high redshift galaxies, the main driver of
the nuclear metallicity shifts to chemical enrichment.  In these
simulations, the nuclear metallicity increases, contrary to what is
observed in the local Universe, suggesting that the interacting galaxy
MZ relation may evolve differently at high redshifts.

We compare our simulations directly to observations by synthesizing a
population of progenitor galaxies with properties consistent with
observed samples and show that the empirical depression in the
close-pair mass-metallicity relation can be reproduced while
accounting for chemical enrichment.  We find good agreement between
our simulations and observations for both the interacting galaxy
mass-metallicity relation and separation-metallicity relation.

%%%%%%%%%%%%%%%%%%%%%%%%%%%%%%%%%%%%%%%%%%%
%%%%%%%%%%%%%%%%%%%%%%%%%%%%%%%%%%%%%%%%%%%
%%%%%%%%%%%%%%%%%%%%%%%%%%%%%%%%%%%%%%%%%%%

%%%%%%%%%%%%%%%%%%%%%%%%%%%%%%%%%%%%%%%%%%%
%%%%%%%%%%%%%%%%%%%%%%%%%%%%%%%%%%%%%%%%%%%
%%%%%%%%%%%%%%%%%%%%%%%%%%%%%%%%%%%%%%%%%%%
\section{Methods}
\label{sec:MergerSimulations}

We employ a library of merger simulations carried out using the
N-body/Smooth Particle Hydrodynamics (SPH) code {\sc
Gadget-2}~\citep{Gadget2}. {\sc Gadget-2} is based on a formulation
of SPH~\citep{SHEntropy} which conserves both energy and entropy
simultaneously (when appropriate).  In addition to accounting for
gravity and hydrodynamics, our simulations also include a sub-grid two-phase
model of star formation, supernova feedback, radiative cooling of gas,
and star formation driven winds~\citep[][hereafter
SH03]{KatzCooling,SHMultiPhase}.  These features give a working
description of the multiphase nature of the interstellar medium
\citep[e.g.][]{MO77} without explicitly resolving the various phases.
Supermassive black hole sink particles are included in our simulations
and have both accretion and thermal feedback associated with them, but
their presence does not influence any results presented in this paper.

The simulations presented here also include a novel method for returning gas to the interstellar
medium (ISM) from (prompt) supernovae and asymptotic giant branch (AGB) winds.
While the main body of our simulations code has been used for numerous
other studies~\citep[see, e.g.,][]{Cox2006b,Cox2006,Robert06,Robert06a,Robert06b},
the inclusion of time--delayed gas recycling is a new feature which is
described.

%%%%%%%%%%%%%%%%%%%%%%%%%%%%%%%%%%%%%%%%%%%
%%%%%%%%%%%%%%%%%%%%%%%%%%%%%%%%%%%%%%%%%%%
%%%%%%%%%%%%%%%%%%%%%%%%%%%%%%%%%%%%%%%%%%%

\subsection{The Basic Model: SH03}
\label{sec:SH03outline}

The primary astrophysical processes modeled in our simulations -- including
cooling, star formation, feedback, and galactic winds -- are included as in the
SH03 model.  Because these processes are discussed extensively in that
paper we review them only briefly here.

The instantaneous rate of star formation for an SPH particle is given by
\begin{equation}
\frac{dM_{*}}{dt} = \frac{M _{c}}{t_{*}}
\label{eqn:sfr}
\end{equation}
where $M _{*}$ is the stellar mass, $M _{c}$ is the mass of cold gas
(i.e. some fraction of the SPH particle's mass that is in the cold
phase as determined via the sub-grid model prescription of SH03), and
$t_{*}$ is a characteristic star formation timescale.  We adopt a 
value $t_* = 4.5$ Gyrs so that our simulations are consistent
with the Kennicutt-Schmidt relation \citep{KS1,KS2,Cox2006}.  Star 
formation is assumed to only take place in gas which has densities
above a set threshold, which in our case is 0.5~cm$^{-3}$.

Equation~\ref{eqn:sfr} is used as the basis for a Monte Carlo method
for actually converting SPH particles to star particles.  Each SPH particle is assigned a
probability of turning into a star particle as follows:
\begin{equation}
\label{eqn:Psfr}
p_* = \left[ 1 - \exp \left( -  \frac{\Delta t}{t _{\mathrm{SFR}}} \right) \right]
\end{equation}
where $\Delta t$ is the current simulation time step, and $t_{\mathrm{SFR}} = M_{gas} / (dM_* / dt)$ is the star formation timescale.
A particle is converted if a random number drawn between 0 and 1 is less than Eq.~\ref{eqn:Psfr}.

The instantaneous rate of mass launched in star formation driven winds is given by 
\begin{equation}
\frac{dM_{w}}{dt} = \gamma \frac{dM _{*}}{dt}
\label{eqn:winds}
\end{equation}
where $dM _{*}/dt$ is the local star formation rate and $\gamma$ is
the mass entrainment efficiency.  We typically take $\gamma=0.3$,
which is the mean value for luminous infrared
galaxies measured by~\citet{RupkeOutflows2005}, and the resulting 
winds are ejected at $v_{\rm w}= 242$~km/sec.

Equation~(\ref{eqn:winds}) is used as the basis for a Monte Carlo method
for wind generation.  The associated probability of launching a
particle in a wind is given by
\begin{equation}
p_{w} = \left[ 1 - \exp \left( - \frac{\Delta t}{t_{w}} \right) \right]
\end{equation}
where $t_w = M_{gas} / (dM_w /dt)$.  Particles put into winds
receive a velocity kick and are not allowed to interact with the
surrounding medium while they leave their host environment.  In
principle, gas particles launched by winds may still contribute to the
nuclear gas phase metallicity.  However, in practice, these particles
very quickly leave the nuclear regions, and no longer contribute to the
nuclear metallicity.

%%%%%%%%%%%%%%%%%%%%%%%%%%%%%%%%%%%%%%%%%%%
%%%%%%%%%%%%%%%%%%%%%%%%%%%%%%%%%%%%%%%%%%%
%%%%%%%%%%%%%%%%%%%%%%%%%%%%%%%%%%%%%%%%%%%

\subsection{Gas Recycling}
\label{sec:GasRecycling}

While models for star formation and feedback are ubiquitous in the current
generation of numerical simulations, it is much less common to include the
recycling of mass and metals which are lost from evolving stellar populations, a fundamental process
in the enrichment of the interstellar and intergalactic mediums.  The few models
that do include this ``gas recycling'' typically do so in a continuous
fashion, during which, at every time step, a small fraction of mass and
metals is transferred from a single stellar particle to a set of nearest
SPH neighbors~\citep[e.g.][]{SteinmetzGasReturn,KawataGasReturn,
KawataGibsonGasReturn,OkamotoSGR,ScannapiecoGasReturn,StinsonGasReturn,TornatoreGasReturn,
WiersmaGasReturn,DilutionPeak2010,PerezScoop}.  This approach can
accurately track the spatial and temporal distribution of mass and metals
at a reasonable computational cost, but unfortunately, because particles can
exchange mass with many neighbors, at many different times, it is impossible to 
trace the exact origin of any particular mass or metal element.

Given a desire to unambiguously track the origin and evolution of interstellar
metals, the work presented here adopts a stochastic approach to gas 
recycling~\citep[e.g.][]{LiaSGR,MartinezSerranoSGR}.
This less-used approach is designed to mirror the stochastic star formation scheme that is typically
implemented in numerical simulations.  While stochastic star formation probabilistically
converts SPH particles into a fixed number of collisionless stellar particles
based upon the star-formation rate, our stochastic gas recycling converts
stellar particles into SPH particles based upon theoretical mass return rates
for a simple stellar population.  When used in conjunction, these methods can
accurately track the temporal and spatial interchange of baryonic material
between stellar populations and the interstellar medium, a necessary element
to any study of the evolution of interstellar metals.

\begin{figure*}
  \begin{center}
   \plottwo{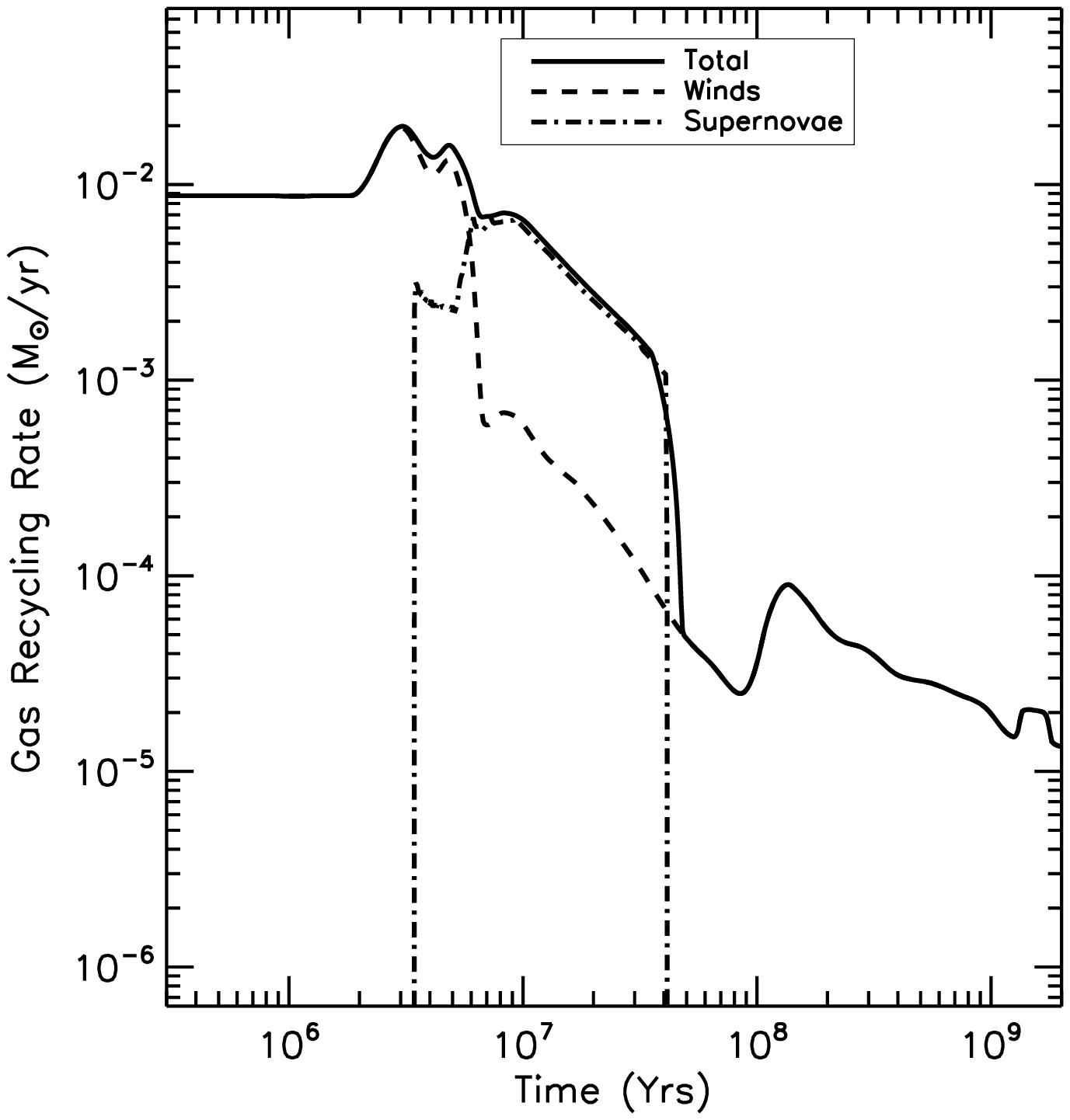}{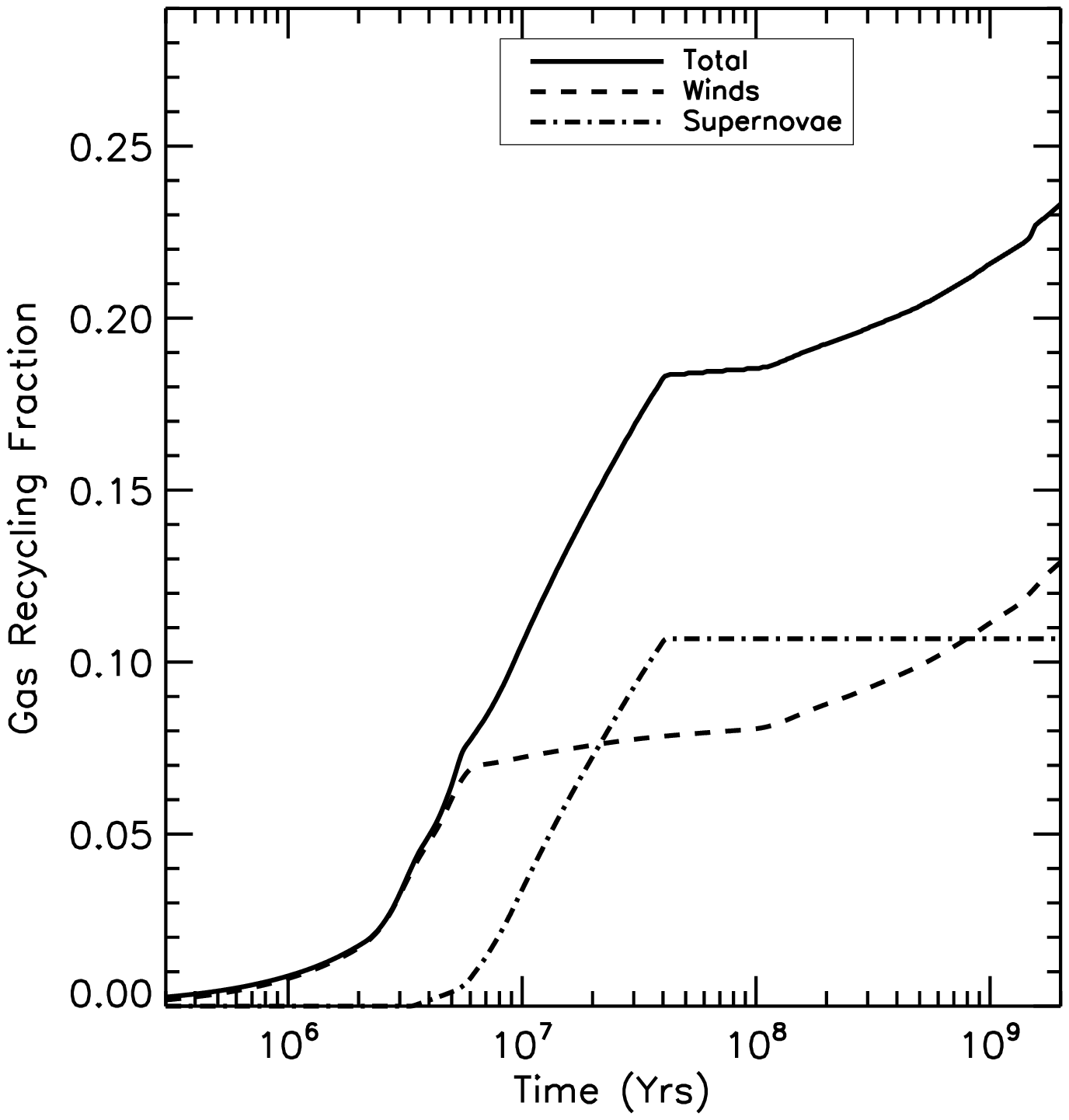}
    \caption{The mass recycling rate (left) and integrated mass recycling fraction (right) from a
    Starburst99 model for a $10^6 M_{\odot}$ stellar population with a Kroupa IMF.}
    \label{fig:massReturnRate}
  \end{center}
\end{figure*}

In our stochastic approach, the probability of turning an individual star particle
into an SPH particle at any given time step is given by
\begin{equation}
p_{recy} = \left[ 1 - \exp \left( - \frac{\Delta t}{t_{recy}} \right) \right]
\label{eq:pgas}
\end{equation}
where $\Delta t$ is the current simulation time step, and
\begin{equation}
t_{recy} = \frac{ M_* (t) }{ dM_{gas}/dt }.
\label{eq:tgas}
\end{equation}
Here, $M_*(t) $ is the expected stellar mass at time $t$, 
and $dM_{gas}/dt$ is the mass return rate at the same time.  
While the physical mass of the stellar particle in the 
simulation remains constant, the expected stellar mass 
identifies the amount of mass that would have been returned 
to the ISM if we implemented continuous gas recycling.  We should make 
clear that there are no hybrid particles in the simulation, so the expected stellar 
mass has no influence on the simulations dynamics.
It is necessary, however, for Eq.~\ref{eq:tgas} to use the expected stellar mass
in order for the stochastic recycling routine to reproduce the 
desired Starburst99 gas recycling rate in the simulations.

In practice, the simulation selects a random number between 0 and 1, and performs the
particle conversion if this number is less than $p_{recy}$.  We calculate the mass return
rate, $dM_{gas}/dt$, directly from Starburst99 synthetic
population models~\citep{Starburst99_1,Starburst99_2,Starburst99_3}.  These
models include gas recycling contributions from core collapse supernova and
asymptotic giant branch (AGB) winds, but do not include contributions from
Type Ia supernovae.  At each time step, stellar particles are assigned a mass
return rate determined by their age, metallicity, and an initial mass function ($\Upsilon$).
For the simulations presented here, we choose a Kroupa initial mass function
with solar metallicity.  The resulting mass return rate, and mass return fraction
are demonstrated in Figure~\ref{fig:massReturnRate}.

While it will be fruitful to explore variations in the mass return according to varying
IMF assumptions, real--time metallicity information, or even easily parameterized
functional forms of mass return that include Type Ia supernovae ~\citep[see, e.g.,][]{JungwiertGasReturn,Kravtsov}
we choose to allow variations in mass return through a simple scale factor to
the return rate shown in Figure~\ref{fig:massReturnRate}.  Specifically, we
define a scalar, $\zeta$, and employ a mass return rate normalized as such
\begin{equation}
\frac{dM_{gas}}{dt} = \left( \frac{ \zeta }{\frac{1}{M_*}\int _0 ^\infty f_{sb99}(t , Z, \Upsilon) dt } \right) f_{sb99}(t , Z, \Upsilon),
\label{eqn:gas_recycling_rate}
\end{equation}
where $f_{sb99}(t , Z, \Upsilon)$ is the mass return rate from Starburst99 as shown
in Figure~\ref{fig:massReturnRate}.  The expected stellar mass in Eq.~\ref{eq:tgas} is
then calculated as
\begin{equation}
M_* (t) = M_0 - \int_0 ^{t } \frac{dM_{gas}}{dt'} dt',
\end{equation}
where $M_0$ is the original mass of the stellar particle.  We note that for large
values of $\zeta$, and for old stellar ages, $M_*(t)$ can become negative, yielding
unphysical negative values for $t_{recy}$ and consequently a negative probability
$p_{gas}$.  This situation is non-catastrophic, however, because, in practice, it
leads to an immediate conversion of this stellar particle to SPH.

The parameter $\zeta$ allows us to control the timescale and efficiency of gas recycling.
For example, the Kroupa IMF we employ dictates that 37\% of the stellar mass will be
returned to the ISM after 10 Gyr.  Setting $\zeta=0.37$ makes the normalization unity
and the recycling therefore tracks exactly what is shown in Fig.~\ref{fig:massReturnRate}.
Setting $\zeta=1$ will produce complete (i.e., 100\%) recycling of the stellar mass
within 10~Gyr, while $\zeta=100$ effectively yields the instantaneous recycling approximation
with all the stellar mass being returned within $\sim3\times10^6$~Yr.  Finally, and
perhaps obviously, setting $\zeta=0$ turns off recycling altogether giving us the 
flexibility to study the impacts of varying amounts of recycling on metallicity evolution.

When a star particle is returned to the gas phase,
the particle type is instantly converted from stellar to gas at the
end of the current time step.  The
fields which are defined for both SPH particles and star particles,
such as the position, velocity, mass, metallicity, etc., remain
unchanged during the conversion.  For concreteness, the metallicity of a
particle remains unchanged during the transition to or from the stellar
state.  The metallicity increases while in the gas phase (according to
Eq~\ref{eqn:MetalEnrichment}), but remains fixed while in -- or making
transitions to and from -- the stellar state.

SPH quantities must be initialized for the newly converted gas particle.
In principle, we could set these fields to
properly reflect the physical state of the gas being returned and
explore the feedback implications that naturally result from this gas
recycling model.  However, in practice, we are resolution-limited, we
already include feedback according to the SH03 model, and our primary
concern is representing the overall mass-budget faithfully.  To this
end, we give a newly formed SPH particle properties that will allow it
to quickly homogenize into the surrounding medium.  In particular, the
entropy is set equal to that of gas at a temperature of $T = 50,000$
K and a density (chosen to be the star formation critical density)
of $\rho = 0.5 \;\mathrm{cm ^{-3}}$.  It is important to note that
our simulations are nearly invariant to our choice of these values.
Changing the initial entropy by an order of magnitude in either
direction yields no obvious differences in the simulations.  In
effect, this implementation of gas return serves to provide a passive
source of gas for our evolving system, without providing strong
feedback.

During both star formation and gas recycling events, no particle
splitting occurs.  As such, the total number of baryon particles
(stars and gas) is conserved throughout the simulation.  Effectively,
baryon particles are permitted to flip back and forth between the gas
and stellar phases, according to the star formation and gas recycling
rates.  Because baryon particle number is conserved and no mass is transferred 
between particles, mapping a particle's initial position to a final position 
a trivial task, regardless of the number of times it was it was turned into 
a star particle or returned to the gas phase.  Using this stochastic gas recycling
method provides the distinct advantage that we 
can retrace all mass to a unique initial position 
regardless of the star formation or gas recycling history.

%%%%%%%%%%%%%%%%%%%%%%%%%%%%%%%%%%%%%%%%%%%
%%%%%%%%%%%%%%%%%%%%%%%%%%%%%%%%%%%%%%%%%%%
%%%%%%%%%%%%%%%%%%%%%%%%%%%%%%%%%%%%%%%%%%%

\subsection{Metallicity Enrichment}
\label{sec:MetallicityEnrichment}
We implement a method for calculating the metal enrichment based on
the instantaneous star formation rate to determine the metallicity
enrichment rate.  Unlike star formation, galactic winds, and gas
return, metal enrichment is carried out in a continuous fashion, where
the metal formation rate is given by
\begin{equation}
\frac{dM_{Z}}{dt} = y \frac{dM_{*}}{dt}
\label{eqn:MetalEnrichment}
\end{equation}
and $y$ is the metal yield and $dM_{*}/dt$ is the instantaneous star formation
rate.  For our simulations, we use a fixed yield of $y=0.02$.  The metallicity will 
increase wherever there is ongoing star formation, as determined via the SH03 star 
formation model.  As such, diffuse gas will not be star forming, and will therefore not 
enrich (see SH03 for details).  The 
scalar metallicity is updated at each time step using an Eulerian integration scheme. 
This scalar metallicity does not track independent species, but instead provides a 
single metallicity value proportional to the integrated star formation rate.  This metal 
enrichment scheme is strictly independent of our gas recycling routine.

The metallicity enrichment does not affect the evolution of the
simulation because there are no metallicity dependent dynamical
processes included, such as metal line cooling.  Because no mixing is
allowed, nearby particles may have large metallicity variations.
This is not problematic, as long as we interpret the metallicity at a
given location to be a kernel weighted average of nearby particles, as
is traditional for determining fluid quantities in SPH.  Mixing
would homogenize the individual particle metallicities, while leaving
the average value of the kernel weighted metallicity unchanged.

Using this enrichment scheme, we are able to cleanly decompose the
metallicity of an SPH particle into contributions from its initial
metallicity and star formation history.  More important, we are able
to scale our metal yield and modify our initial metallicity setup in
our post-processing analysis, without requiring additional
simulations.  This is possible because all of the gas content of
an SPH particle has an unambiguous and unique initial location and
star formation history, which is not true when particle mass mixing is used.  
In later sections, our ability to arbitrarily
modify initial metallicity gradients of our progenitor galaxies is
critical in allowing us to thoroughly sample the range of metallicity
gradients of our progenitor galaxies, without large computational
requirements.

%%%%%%%%%%%%%%%%%%%%%%%%%%%%%%%%%%%%%%%%%%%
%%%%%%%%%%%%%%%%%%%%%%%%%%%%%%%%%%%%%%%%%%%
%%%%%%%%%%%%%%%%%%%%%%%%%%%%%%%%%%%%%%%%%%%

\section{Isolated Galaxies}
\label{sec:IsolatedGalaxyMetallicityEvolution}

In our analysis, we take the nuclear region of a galaxy to be a
sphere of radius 1 kpc about the galactic center.  We determine the nuclear metallicity
from the star formation rate weighted average of all gas particles inside this 
sphere.  A 1 kpc spherical region is used for consistency with the 
observations of~\citet{KGB06}.  However, it should also be noted that our results 
would not fundamentally change for slightly larger or smaller definitions of the nuclear region.  

We use star formation rate weighted averages to mimic
observations of HII regions, which naturally select star forming gas.
Unless otherwise stated, all nuclear metallicities quoted in this
paper are gas-phase and star formation rate weighted.  In particular,
the central depressions in gas-phase metallicity presented here should
not be confused with enhancements in stellar metallicity seen in
late-stage mergers and relaxed elliptical galaxies.  These increases
in stellar metallicity are relics from the merger-driven starbursts
that leave behind central stellar cusps in merger remnants
\citep{MH94b,Hopkins08a} and elliptical galaxies
\citep{Hopkins09a,Hopkins09b}.  These starbursts occur at late
stages in a merger, following coalescence, from enriched, star-forming
gas, yielding a central population of young, metal-enhanced 
stars \citep[see, e.g. Figs. 27 and 28 in ][]{Hopkins09a}.

The star formation rate weighted nuclear metallicity is defined as
\begin{equation}
\bar{Z} =  \frac{\int _V \dot \rho_* \left( \vec r \right) Z \left( \vec r \right) dV}{\int_V  \dot \rho_* \left(  \vec r \right) dV }
\end{equation}
where $\dot \rho_* \left( x,y,z \right)$ is the star formation rate
density, $Z \left(x,y,z \right)$ is the fraction of gas-mass in
metals, and the integral is performed over the 1 kpc spherical region
about the galaxy's center.  The star formation rate density is calculated as
\begin{equation}
\dot \rho_* \left( \vec r \right)  = \sum _j \frac{dM_{*,j}}{dt} W \left( \left| \vec r - \vec r _j \right| , h_j \right)
\end{equation}
where $W \left( \left| \vec r - \vec r _j \right| , h_j \right)$ is a smoothing kernel function, $h$ is 
a smoothing length, and
$dM_{*}/dt$ is a particle's instantaneous star formation rate (discussed in 
\S\ref{sec:SH03outline}).  For our analysis, we use a galaxy's central black hole to define the galactic 
center; however, our results are unchanged if we use the potential minimum, 
or any other reasonable measure of the galactic center.  This approach neglects projection
effects, but instead provides information about the ``true'' nuclear
region.

The metallicity of each SPH particle consists of two separate
contributions: an initial metallicity and an enriched metallicity.
The initial metallicity of an SPH particle is completely determined by
the particle's initial radial position (discussed in
\S\ref{sec:GalaxySetup}) and is unchanging in time, while the enriched
metallicity grows owing to star formation.  The metallicity of an SPH
particle as a function of time is given by
\begin{equation}
\label{eqn:z_t}
Z_{gas}\left(t \right) = Z_{init} + \frac{y}{M_{gas}} \int_{0} ^{t} \frac{d M_{*}(t')}{dt'} dt'
\end{equation}
where $d M_{*} (t')/dt' $ is the particle's complete time dependent
star formation rate history.  The integration of the rightmost term in Eq.~\ref{eqn:z_t} should be carried out
starting at the beginning of the simulation (i.e. $t=0$), regardless of whether a particle has been recycled or not.  Since 
$dM_{*} (t') /dt'$ is non-zero only when a particle is actively star forming (i.e. in the gas state 
and above the star formation threshold density), periods of time 
when a particle is in the stellar state or below 
the star formation threshold density (see \S\ref{sec:SH03outline}) will not contribute to Eq.~\ref{eqn:z_t}.  

%Given a particle's complete star formation rate history -- which 
%is tracked through the simulation -- we can pick a value for the particle's initial metallicity $Z_{init}$ 
%and immediately determine that particle's resulting metallicity at any arbitrary time.

The metallicity of the nuclear region is calculated similarly to the metallicity of an individual SPH particle, with an additional sum over an ensemble of particles weighted by their star formation rates.  We explicitly break the nuclear metallicity into two components
\begin{equation}
\label{eqn:z_components}
\bar Z = \frac{\sum_{i} Z_{i,init} \dot M_{*,i}}{\sum_{i} \dot M_{*,i}} + \frac{\sum_i Z_{i,enrich} \dot M_{*,i}}{\sum_i  \dot M_{*,i}}
\end{equation}
where the sum is performed over the particles that fall within the
nuclear region.  The first term contains the contribution
from the initial metallicity gradient of our progenitor
galaxies while the second term contains the contribution from star
formation induced metallicity enrichment.  For clarity throughout, we
call the first term the ``dynamical metallicity'', as it is a
dynamical result of the initially assumed metallicity gradient, while
we call the second term the ``enriched metallicity'', as it is a
product of metal enrichment.  For concreteness, the metallicity
gradients for one of our ``C'' isolated galaxies evolved in isolation for 2 Gyrs
is shown in Figure~\ref{fig:isoGal} with contributions from the
dynamical and enriched components explicitly shown (see
\S\ref{sec:IsoGalEvo}).

\begin{table*}
\begin{center}
\caption{Progenitor Disk Properties}
\label{MergerProperties}
\begin{tabular}{ c c c c c c c c }

& & & & & & & \\
Disk Identifier &Total Halo & Initial Stellar & Initial Disk   & $h$ & $N_{Halo}$ & $N_{gas}$ & $N_{stars}$ \\
& Mass $ [M_{\odot}]$ &  Mass $[M_{\odot}]$   &   Gas Fraction & [kpc] & & & \\
 & & & & & & & \\
\hline
\hline 
 & & & & & & &\\
A  & $2.3 \times 10^{11}$  &  $1.1 \times 10^{10}$ & $10\%$ & 2.4 & 532,500 & 30,000 & 145,500 \\
B & $5.1 \times 10^{11}$  &  $2.4 \times 10^{10}$ & $10\%$  & 3.2 & 532,500 & 30,000 & 145,500 \\
C & $9.5 \times 10^{11}$  &  $4.4 \times 10^{10}$ & $10\%$   & 3.9 & 532,500 & 30,000 & 145,500\\
D & $13.5 \times 10^{11}$  &  $6.2 \times 10^{10}$ & $10\%$  & 4.4 & 532,500 & 30,000 & 145,500\\
& & & & & & &\\
C2 & $9.5 \times 10^{11}$  &  $4.0 \times 10^{10}$ & $20\%$   & 3.9 & 532,500 & 40,000 & 135,500 \\
C3 & $9.5 \times 10^{11}$  &  $ 3.2 \times 10^{10}$ & $ 40\% $   & 3.9 & 532,500 & 60,000 & 115,500\\
C4 & $9.5 \times 10^{11}$  &  $2.4 \times 10^{10}$ & $60\%$   & 3.9 & 532,500 &80,000 &95,500\\
& & & & & & &\\
\hline
\hline
\end{tabular}
\end{center}
\end{table*}

%%%%%%%%%%%%%%%%%%%%%%%%%%%%%%%%%%%%%%%%%%%
%%%%%%%%%%%%%%%%%%%%%%%%%%%%%%%%%%%%%%%%%%%
%%%%%%%%%%%%%%%%%%%%%%%%%%%%%%%%%%%%%%%%%%%

\subsection{Isolated Galaxy Setup}
\label{sec:GalaxySetup}

The isolated galaxies used in this paper are modeled following the
analytical work of \citet{MMWdisk} employing the procedure outlined
in~\citet{SDMHModel}.  Our fiducial galaxy consists of a dark matter
halo, an embedded rotationally supported exponential disk, a stellar
bulge, and a central supermassive black hole.

We construct a set of four isolated galaxies with total system masses
ranging from~$\sim 10^{11} M_{\odot}$ to $\sim 10^{12} M_{\odot}$, as
outlined in Table~\ref{MergerProperties}.  Although varied in mass, all 
systems are constructed to be self-similar in order to isolate the effects of mass
from other quantities that may correlate with mass in observed systems.  The total 
mass of the disk (stars and gas combined) is chosen to be a constant fraction (4.0\%)
of the halo mass for all progenitor galaxies, with four settings for the initial
gas fraction (8\%, 20\%, 40\%, and 60\%).  All halos have spin 
parameters equal to $\lambda =0.05$,
which effectively sets the disk radial scale length.  The initial stellar disk scale height is 
set to a fixed fraction (0.2) of the initial disk scale length, while the gaseous scale height is 
determined by satisfying hydrostatic equilibrium~\citep{SDMHModel}.   Stellar bulges
are included as part of the fiducial galaxy setup and are given a fixed fraction (1\%) of
the system mass.  The resulting disk setup has a Toomre Q parameter that varies as a function 
of galactocentric radius, but is everywhere greater than unity -- indicating stability against axisymmetric perturbations.

The dark matter halos follow ~\citet{H90} profiles with concentration
indices of $c=10$.  The stellar and gaseous disk components are modeled
with exponential surface density profiles
\begin{equation}
\Sigma _{*,g} \left( r\right) = \frac{M_{*,g}}{2 \pi h^2} \exp \left( -\frac{r}{h} \right)
\label{eqn:sigma}
\end{equation}
where $M_{*,g}$ is the total mass of stars and gas in the disk, respectively,
and $h$ is their common scale length set by the disk's angular momentum
~\citep{MMWdisk, SpringelWhiteDisk}.  Stellar bulges are taken to be spherical
~\citet{H90} profiles, where the bulge scale length is given in terms of the
halo scale length, $a_b = 0.2 a$, and the bulge mass fraction to be 1\% of the
halo mass.

The initial metallicity profile of our progenitor galaxies enters as an
assumption.  We model the metallicity profile of the disk as an exponential
\begin{equation}
Z \left( r \right) = Z_0 \; \exp \left( - \frac{r }{ h_z  } \right)
\label{eqn:initZ}
\end{equation} 
where the central metallicity, $Z_0$, and the metallicity scale length, $h_z$,
are chosen to be consistent with observations. Isolated galaxies show 
dispersion in their radial metallicity gradients.  We could include this dispersion in 
our simulations by adding in a Gaussian error term to Eq.~\ref{eqn:initZ}; However, 
no dispersion is used when setting 
the initial metallicity profile for our isolated galaxies because this would add 
an extra element of uncertainty into our models and add 
noise to our results, without yielding systematic changes or further insight.

We assume that Oxygen makes up 30\% of the metals by mass.  The constant
Oxygen mass fraction is a necessary assumption of our models, because we track
only one scalar metallicity value.  We note that our results are not particularly sensitive
to our choice of the Oxygen mass fraction, since this is merely a normalization of the
overall metallicity.

We fix the central metallicity values to the observed relation of \citet{Tremonti2004} so
that the central metallicity is given in terms of the stellar mass
\begin{equation} 
12+\mathrm{log} \left( O/H\right) = -1.492 + 1.847 x - 0.08026 x^2
\label{eqn:MZTremonti}
\end{equation}
where $x=\mathrm{log}_{10} M_*$.  Initially, we use the exact value given by
equation~(\ref{eqn:MZTremonti}) when discussing the process of nuclear metallicity
evolution in \S\ref{sec:MergerSetup}.  However, in \S\ref{sec:ComparisonToObservations}
we pick the central metallicity of each progenitor galaxy from a Gaussian distribution
with a standard deviation, $ \sigma = 0.1$ dex, taken from \citet{Tremonti2004}.  Similarly, we use 
the average result from \citet{Z94} of $h_z =  h/0.2$ in \S\ref{sec:MergerSetup}.  However, 
in \S\ref{sec:ComparisonToObservations} we pick the slope of the metallicity gradient from 
a Gaussian distribution with a standard deviation of $\sigma = 0.3 h_z$.

%%%%%%%%%%%%%%%%%%%%%%%%%%%%%%%%%%%%%%%%%%%
%%%%%%%%%%%%%%%%%%%%%%%%%%%%%%%%%%%%%%%%%%%
%%%%%%%%%%%%%%%%%%%%%%%%%%%%%%%%%%%%%%%%%%%

\subsection{Isolated Galaxy Evolution}
\label{sec:IsoGalEvo}

While the metallicity evolution of isolated galaxies is not the focal point
of this paper, we take a moment in this section to demonstrate that our
galaxy models are indeed stable, and evolve very little in isolation.  We
specifically demonstrate that the galaxy models do not develop strong bars
since these structures can lead to significant radial inflows of gas.  Such
processes obscure the true origin of any central metallicity evolution, and
what is specifically a result of the merger.

\begin{figure*}
  \begin{center}
   \includegraphics[height=2.0in]{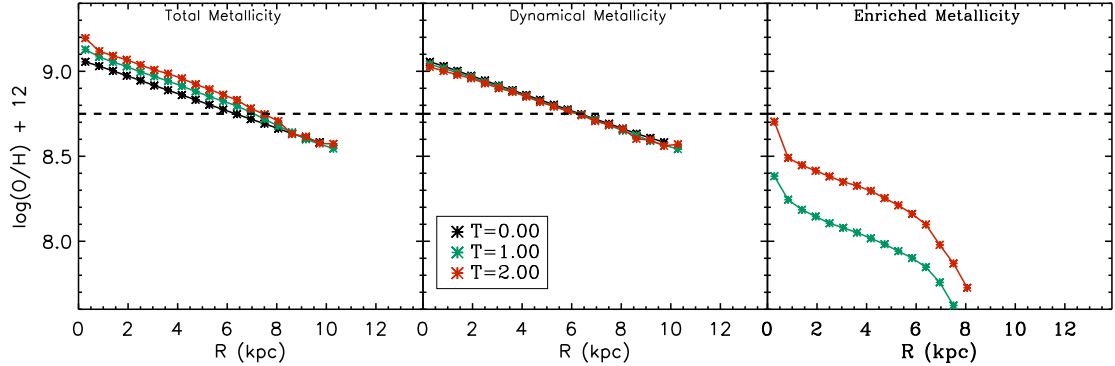}
    \caption{Metallicity profiles for the ``C'' isolated galaxies evolved for 2 Gyr.  The total (left), dynamical (center), and enriched (right) metallicities are shown separately.  The horizontal dashed line denotes the mean initial metallicity for the disk.  The stability of the dynamical metallicity
    indicates that radial gas flows are minimal in our isolated systems, while the total metallicity gradient increases over time 
owing to chemical enrichment.  A clear distinction is made between the dynamical metallicity, which is a product of our initial conditions, from the enriched metallicity, which is a product of ongoing star formation in the simulation.  }
    \label{fig:isoGal}
  \end{center}
\end{figure*}

\begin{figure*}
  \begin{center}
\plotone{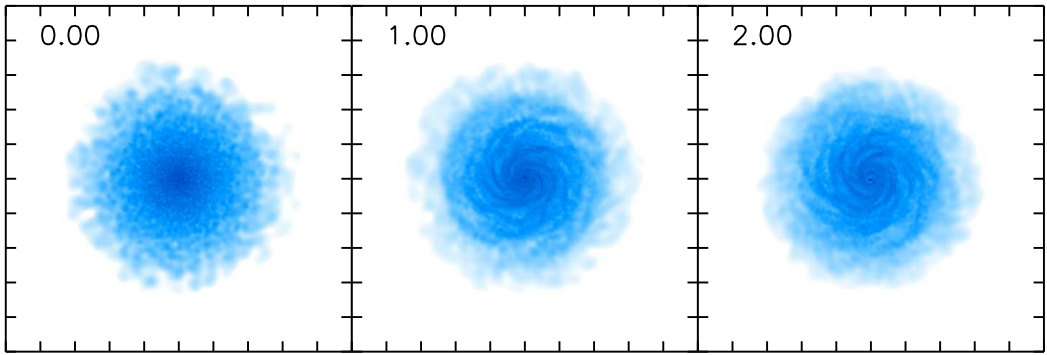}
    \caption{The gas surface density is shown for our isolated ``C'' disk evolved for 2 Gyrs.  The disks evolve stably and without bar formation or significant radial gas inflow for 2 Gyrs.   }
    \label{fig:gasPanel}
  \end{center}
\end{figure*}

\begin{figure*}
  \begin{center}
\plottwo{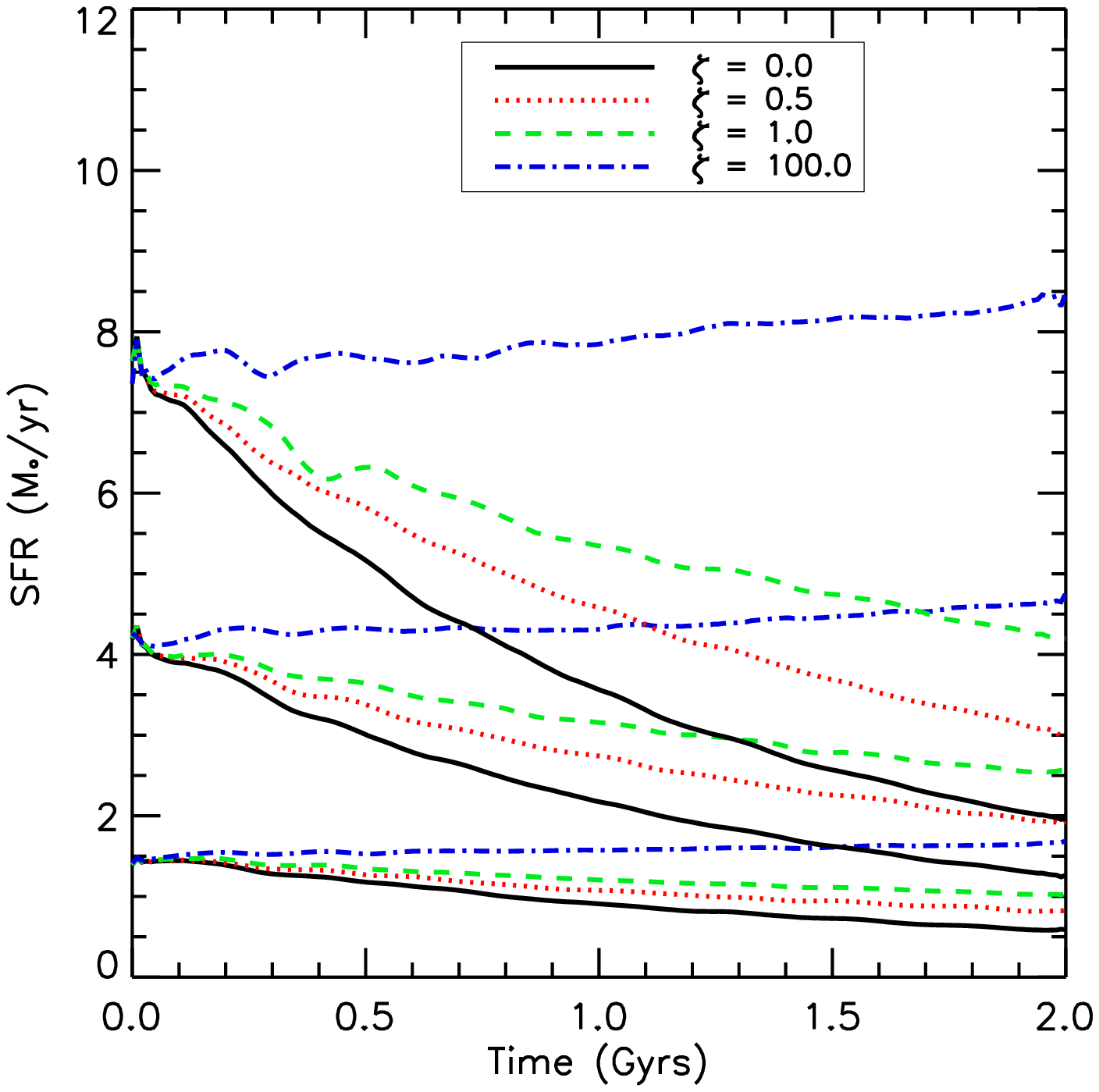}{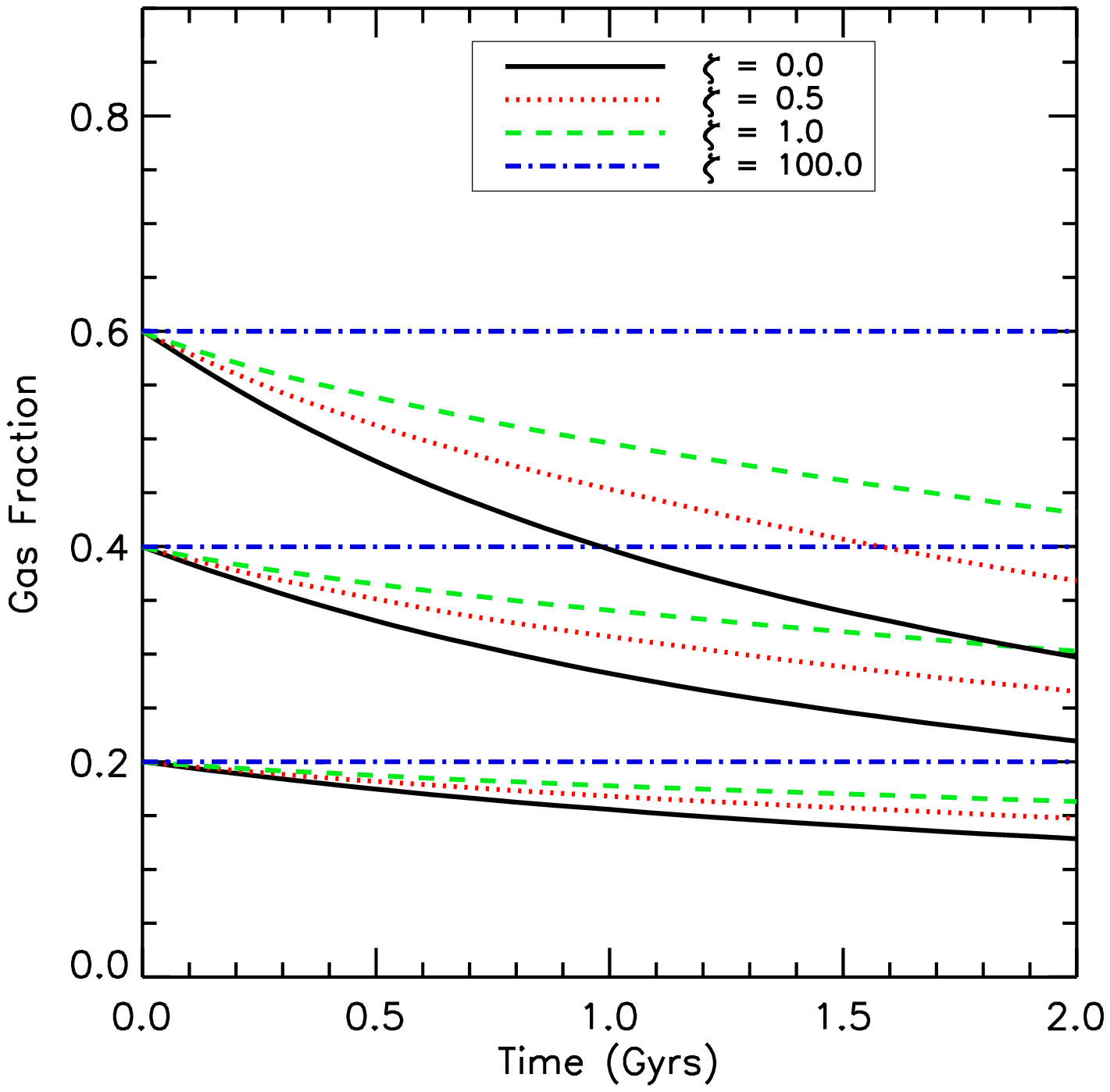}
    \caption{Star formation rates and gas fractions for the C2, C3, and C4 systems evolved for 2 Gyrs with varying gas recycling parameters.  The gas return parameters, identified in the legend, are varied from $\zeta = 0.0$ (i.e. no gas recycling) to $\zeta = 100.0$ (i.e. nearly instantaneous recycling).  Unless otherwise noted, subsequent sections and simulations use $\zeta = 0.3$, which is the unscaled value taken from the Starburst99 simulations.}
    \label{fig:IsolatedDisksExample}
  \end{center}
\end{figure*}

The evolved gas surface density, seen in Figure~\ref{fig:gasPanel},
demonstrates that our disks are both stable and free of strong bars over
at least 2 Gyrs, the typical duration of the galaxy major mergers we study.
As a result, the metallicity gradients of our isolated systems evolve very
little, as seen in Figure~\ref{fig:isoGal}.  We further demonstrate the 
characteristics of our isolated disks in Figure~\ref{fig:IsolatedDisksExample} 
where we show the star formation rates and gas fractions for the C2, C3, and
C4 systems.

To understand where the small amount of metallicity evolution does come from,
Figure~\ref{fig:isoGal} also shows the evolution of the initial metallicity
gradient, or the ``dynamical'' metallicity (middle panel), and the contribution
from ongoing star formation, or the ``enriched'' metallicity (right panel).  From this information it is
easy to see that the small increase in central metallicity and gradient
is a result of chemical enrichment from ongoing star formation, while the initial 
metallicity gradient remains nearly unchanged.

Motivation to emphasize the stability of our initial disk models stems from the
fact that the disks used in previous studies~\citep[e.g.][]{DilutionPeak2010} appear to be bar
unstable over short timescales (see their Figure~2 in their Appendix).  These
bars result in strong radial mixing and a flattening of their radial metallicity
profiles (see Figure~1 in their Appendix).  While bar induced gas inflows are
indeed a valid physical mechanisms for modifying the nuclear metallicity, we
wish to eliminate this complication from our study to make our results easier
to interpret.

\begin{figure}
  \begin{center}
   \plotone{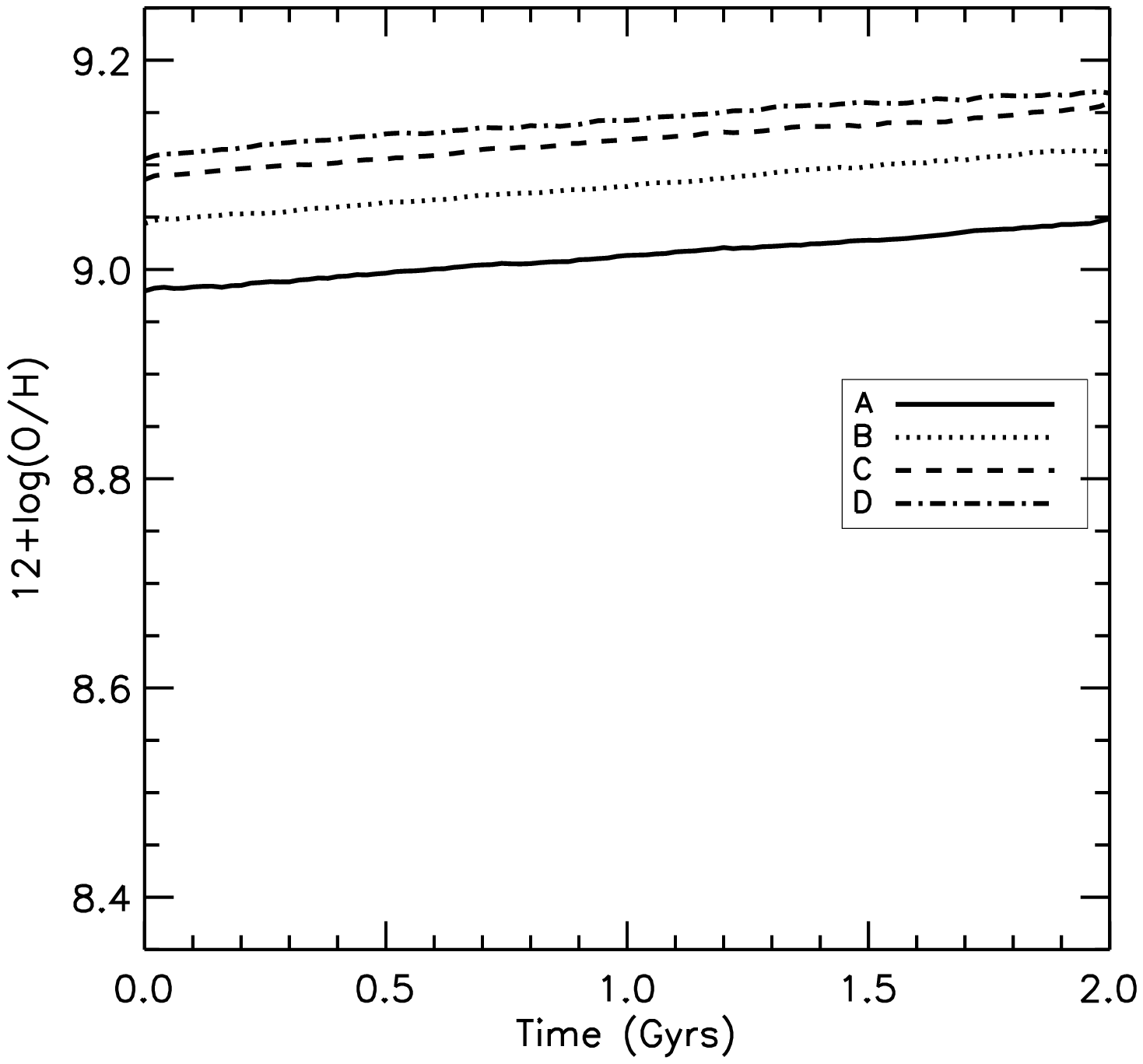}      
    \caption{Nuclear metallicities for the A, B, C, and D, isolated 
    galaxies evolved for 2 Gyrs is shown.  The 
     nuclear metallicity of our isolated model galaxies increases 
     monotonically with time owing to chemical enrichment, as demonstrated
     in Figure~\ref{fig:isoGal}}
    \label{fig:IsolatedDisksStats}
  \end{center}
\end{figure}

The evolution of the nuclear metallicity of the four progenitor disks,
described in Table~\ref{MergerProperties}, are demonstrated in
Figure~\ref{fig:IsolatedDisksStats}.  The nuclear metallicity of all isolated
disks is well described by a mild and monotonic increase over at least
2 Gyrs.  The steady behavior of the isolated galaxy nuclear
metallicity ensures that any strong changes during the merger
simulations are a product of the merger process, rather than disk
instability.

While exploring the behavior of isolated disks, we can also
demonstrate the effects and capabilities of stochastic gas recycling.
When evolving the ``C'' systems (i.e.  C2, C3, and C4 from
Table~\ref{MergerProperties}) we vary the gas recycling parameter and
examine the resulting gas fraction and star formation rate evolution.
The results, shown in Figure~\ref{fig:IsolatedDisksExample},
demonstrate that the gas is locked into stars more efficiently when
the gas recycling parameter is small.  As a result, the star formation
rates, which largely depend on the amount of gas available for star
formation, are consistently larger for increasing values of the gas
recycling parameter.  Simulations that use efficient gas recycling
maintain larger gas fractions and star formation rates throughout the
simulation.

In the limit of very large gas recycling parameters, all newly formed stellar material is quickly 
returned to the ISM.  As a result, the gas fraction for large gas recycling
parameters stays nearly constant.  The corresponding star formation rates slowly 
increase as gas naturally cools and condenses over time.

%%%%%%%%%%%%%%%%%%%%%%%%%%%%%%%%%%%%%%%%%%%
%%%%%%%%%%%%%%%%%%%%%%%%%%%%%%%%%%%%%%%%%%%
%%%%%%%%%%%%%%%%%%%%%%%%%%%%%%%%%%%%%%%%%%%

\section{Merging Galaxies}
\label{sec:MergerSetup}

While keeping all of the conventions for nuclear metallicity defined in 
\S\ref{sec:IsolatedGalaxyMetallicityEvolution}, we now consider the 
evolution of galactic nuclear metallicity for merging systems.  Since 
our interacting systems are spatially extended and overlap substantially during 
the interaction, they do not follow Keplerian trajectories.  However, we use the terminology
of Keplerian orbits to clearly describe the merger setup.  

\begin{figure}
  \begin{center}
    \includegraphics[height=2.7in]{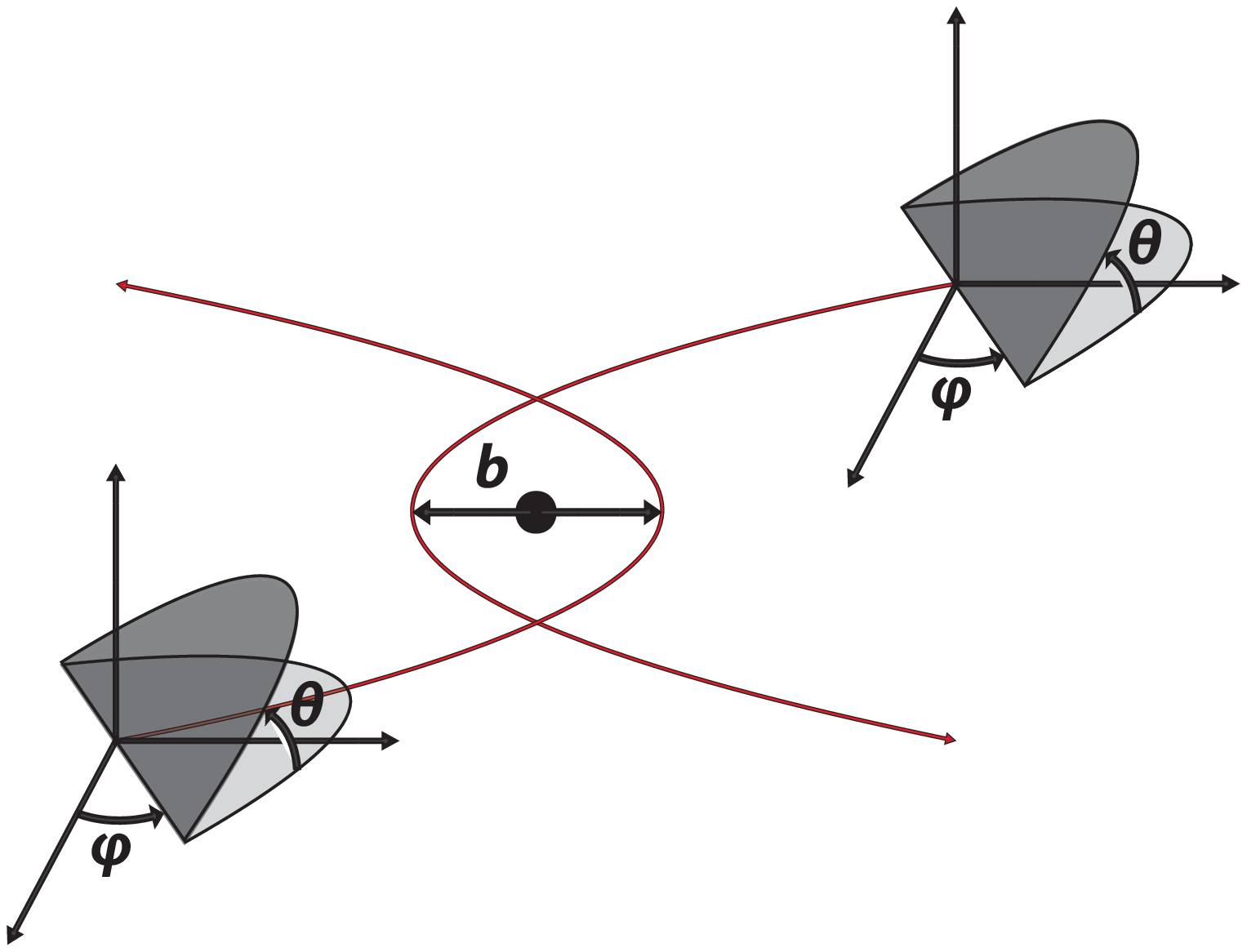}
    \caption{Schematic representation of the merger setup.  Our coordinate system is defined by the plane of the merger orbit, and each disk's orientation is independently adjusted with respect to this plane.}
    \label{fig:diskOrientation}
  \end{center}
\end{figure}

To completely specify our merging setup, we must specify the properties of the 
merger orbit (2 parameters), and the relative orientation of each galaxy with respect 
to the orbit (2 parameter for each galaxy).  We start by specifying that the systems 
will be on zero energy orbits (i.e. eccentricity value of unity for Keplerian objects).  The orbital angular 
momentum is then set by picking a value for the impact parameter, assuming 
Keplerian trajectories (the ``real'' impact parameter found in the simulations
will be larger than this value).  We hold the impact parameter fixed (5 kpc) for 
all simulations.  The spin angular momentum vector of each galaxy is varied with respect to 
the orbital angular momentum vector according to the orientations detailed in 
Table~\ref{OrbitalOrientations} as visually depicted in Figure~\ref{fig:diskOrientation}.

The parameter space of mergers is quite large.  To explore this
parameter space, without using an excessively large number of
simulations, we first perform a detailed exploration of a single case.
We then systematically vary merger parameters, such as the disk
orientation, progenitor disk mass, merger mass ratio, orbital angular
momentum, and so forth.  While this does not necessarily cover all of
merger parameter space, it does enable us to paint a coherent picture
detailing the relation between the evolution of the merger and the
evolution of the nuclear metallicity.

We select our fiducial galaxy merger to be two identical ``B'' disks
merging on the ``e'' trajectory (see Tables~\ref{MergerProperties}
and~\ref{OrbitalOrientations}).  This setup is neither average nor
special, and is simply selected to demonstrate in detail the
relationship between the merger state and evolution of the nuclear
metallicity.  While \S\ref{sec:CaseStudy} is restricted to
understanding a single merger evolution,
\S\ref{sec:MergerOrientation}-\ref{sec:MergerGasFraction} show that
much of the nuclear metallicity evolution can be understood in terms
of generic merger properties that will continue to drive the nuclear
metallicity evolution as we move beyond this fiducial setup.

\begin{table}
\begin{center}
\caption{Orbital Orientations}
\label{OrbitalOrientations}
\begin{tabular}{ c  c c c c  }

Orientation & $\theta _ {1}$ & $\phi _{1}$ & $\theta _{2}$ & $\phi _{2}$ \\
Identifier & [deg] & [deg] & [deg] & [deg]  \\
    & & & &  \\
\hline
\hline
 & & & & \\
a & 90 & 90 & 0 & 0 	\\
b & 180 & 0 & 0 & 0	\\
c & 180 & 0 & 180 & 0 	\\	
d & 90 & 0 & 0 & 0 \\
e & 30 & 60 & -30 & 45  \\
f  & 60 & 60 & 150 & 0 	\\
g & 150 & 0 & -30 & 45 	\\
h & 0 & 0 & 0 & 0 	\\
i & 0 & 0 & 71 & -30 	\\
j & -109 & 90 & 71 & 90	\\
k & -109 & -30 & 71 & -30 	\\
l & -109 & 30 & 180 & 0 	\\	
m & 0 & 0 & 71 & 90	 \\
n & -109 & -30 & 71 & 30 	\\
o & -109 & 30 & 71 & -30 	\\
p & -109 & 90 & 180 & 0 	\\
 & & & &  \\
\hline
\hline
\end{tabular}
\end{center}
\end{table}

%%%%%%%%%%%%%%%%%%%%%%%%%%%%%%%%%%%%%%%%%%%
%%%%%%%%%%%%%%%%%%%%%%%%%%%%%%%%%%%%%%%%%%%
%%%%%%%%%%%%%%%%%%%%%%%%%%%%%%%%%%%%%%%%%%%

\subsection{Metallicity Evolution in Merging Systems}
\label{sec:CaseStudy}

\begin{figure}
  \begin{center}
   \plotone{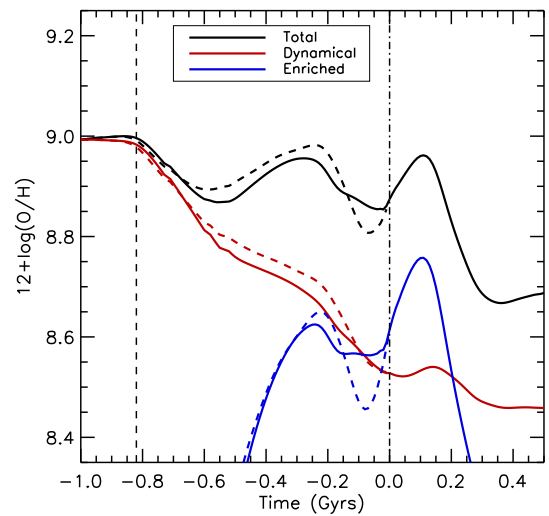}  
\caption{The total (black), dynamical (red), and enriched (blue) nuclear metallicities as a function of time for 
the ``BBe'' merger simulation.  The two disks, which do not have identical metallicity evolutionary tracks, are distinguished by the solid and dashed lines.  First pericenter passage and final coalescence are marked with vertical dashed and dot-dashed lines, respectively.
}
    \label{fig:ZTcasestudy}
  \end{center}
\end{figure}

In stark contrast to the isolated systems presented in
\S\ref{sec:IsoGalEvo}, the nuclear metallicity of interacting galaxies
is a complicated function of time, as shown in Figure~\ref{fig:ZTcasestudy}.  
Moreover, there is a clear change in the evolutionary behavior 
of the nuclear metallicity following pericenter passage when the disks 
stop behaving like isolated systems, and enter a period of 
evolution dominated by the merger dynamics.

Before delving further into the merger details, we note that there is
a characteristic ``double dip'' shape to the nuclear metallicity
evolution.  This is caused by the ongoing competition between
metallicity dilution and chemical enrichment.  While the exact
evolutionary track shown here is specific to this 
particular merger setup, the
characteristic double dip in the metallicity will appear repeatedly in
many other cases.  In general, galactic close passages lead to
gas inflows (yielding inflows of low metallicity gas) which are
quickly followed by star formation (yielding chemical enrichment).

During the period of time between first pericenter passage and final 
coalescence, the nuclear metallicity evolution is driven by: 1) Radial inflows
of low metallicity gas caused by the tidal interaction, 2) Chemical
enrichment caused by star formation activity, 3) Locking of
metals and gas in the stellar phase, and 4) Removal of gas and metals
via galactic winds.  The combined effect of these four mechanisms, 
which are discussed in detail in the following sub-sections,
determines the relative importance of metallicity dilution and 
chemical enrichment, which ultimately decide when and 
where the nuclear metallicity will be diluted or enhanced.

\begin{figure*}
  \begin{center}
   \plotone{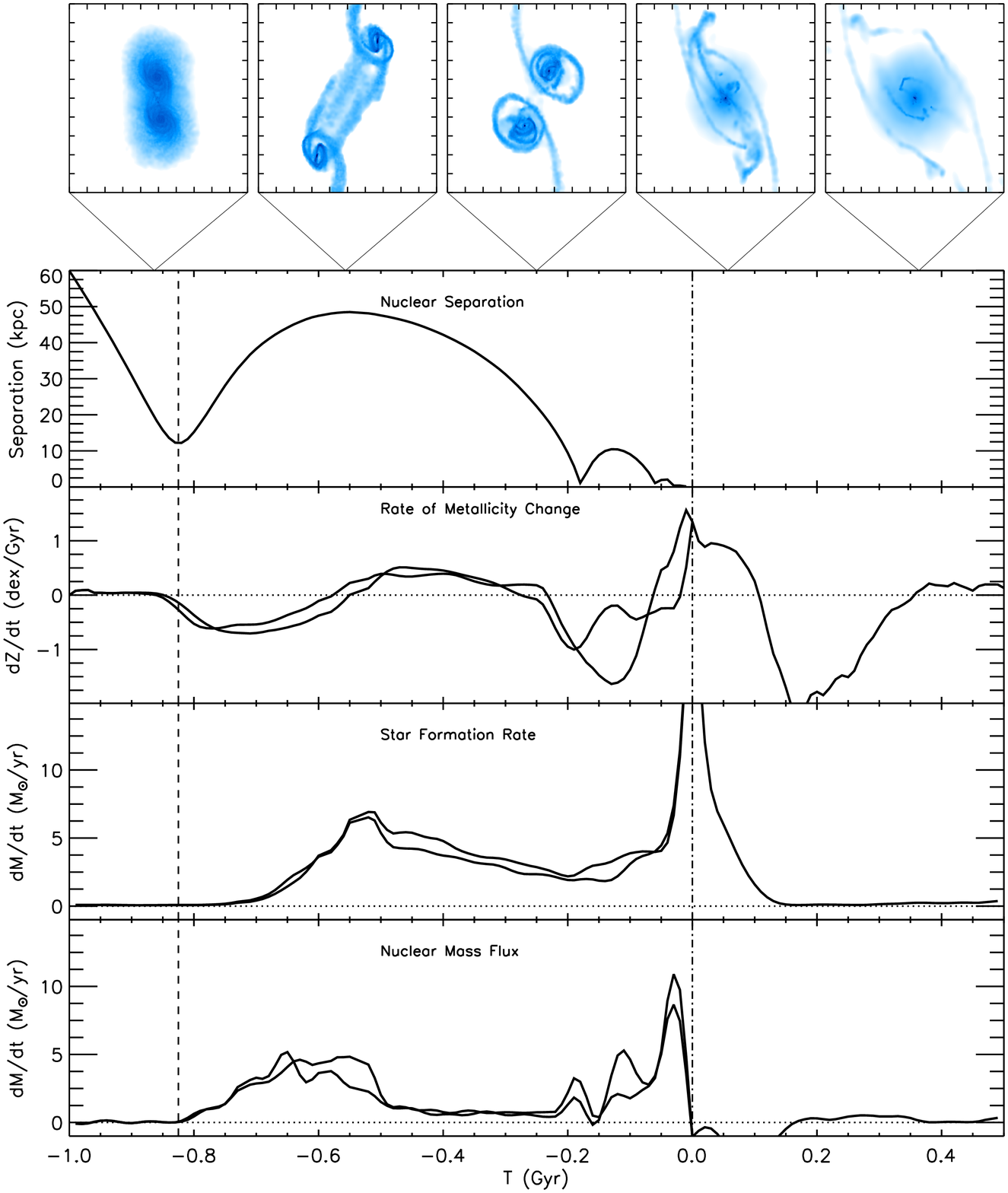}  
\caption{Several diagnostics for assessing the 
metallicity evolution of our fiducial merging 
system are shown.  The top row shows contour 
plots of the gas density, with lines indicating the stage 
of the merger.  From top to bottom, the subsequent time 
series show the galactic nuclear separation, the
rate of change of the nuclear metallicity, nuclear
star formation rate, and nuclear gas inflow rate.  Pericenter
passage and final coalescence are denoted by 
dashed and dot-dashed lines, respectively.  In general,
periods of ongoing nuclear metallicity dilution can be associated 
with strong nuclear gas inflows, while times of ongoing
nuclear metallicity enhancement are associated with 
high nuclear star formation rates.
 }
    \label{fig:ZTCenterPiece}
  \end{center}
\end{figure*}

%%%%%%%%%%%%%%%%%%%%%%%%%%%%%%%%%%%%%%%%%%%
%%%%%%%%%%%%%%%%%%%%%%%%%%%%%%%%%%%%%%%%%%%
%%%%%%%%%%%%%%%%%%%%%%%%%%%%%%%%%%%%%%%%%%%

\subsubsection{Metallicity Dilution and Chemical Enrichment}
\label{sec:DilutionAndEnrichment}
The influence of radial gas inflows and chemical enrichment 
are demonstrated in Figure~\ref{fig:ZTCenterPiece}, which shows 
the galactic separation, rate of change of nuclear metallicity, 
nuclear star formation rate, and nuclear gas inflow rate.  In 
terms of basic merger properties, there is an influx of gas following first 
pericenter passage and periods of gas inflow associated with each 
additional close passage.  These
gas inflows give rise to high nuclear star formation rates.  While 
these previous points have been studied extensively in other papers 
 \citep{BH91,BH96,MH94a,MH96,Iono04}, we instead focus here on the influence that 
these generic merger properties have on the evolution of the nuclear metallicity.
Specifically, times of strong gas inflow correspond to periods of nuclear 
metallicity depression, while high star formation activity aligns with
nuclear metallicity enhancement.  These qualitative relationships 
remain true as the merger parameters are varied.

Previous studies have found that the depression in the nuclear metallicity is 
correlated with the mass of gas that
migrated to the nuclear region~\citep{Rupke2010}.  This result is reproduced in our simulations
when we neglect star formation (similar to the red line in Figure~\ref{fig:ZTcasestudy}) and is 
a clear-cut demonstration of metallicity dilution.  
However, when we also consider contributions 
from chemical enrichment and the consumption
of gas via star formation, this correlation disappears.

Inspection of Figure~\ref{fig:ZTCenterPiece} shows that, in general,
periods of strong gas inflow occur simultaneously with times of
ongoing nuclear metallicity depression.  Similarly, periods of high
star formation give rise to enhancements in the nuclear metallicity.
Interestingly, neither the nuclear star formation rate nor the nuclear
mass inflow rate correlate well with the rate of change of the nuclear
metallicity because they exchange roles playing the dominate driver of the
nuclear metallicity throughout the merger.  The reason is that
metallicity dilution, chemical enrichment, locking of gas in the
stellar phase, and galactic winds all influence metallicity
evolution.  While more involved parameters (e.g. the difference
between the star formation and nuclear mass inflow rate) yield better,
albeit imperfect, correlations, these trends can be misleading
and they tend to over-complicate a fairly simple point.  In
particular, changes to the nuclear metallicity at any time during the
interaction can be understood by examining the nuclear star
formation and gas inflow rates.  Furthermore, the nuclear star
formation and gas inflow rates are naturally explained via the well
studied merger process.

From this single merger example, it seems plausible that the role of
metallicity dilution and chemical enrichment are natural consequences
of the merger process.  However, by inspection of
Figure~\ref{fig:ZTcasestudy}, neither effect is overwhelmingly
dominant.  Hence, in the following analysis we pay attention not
only to metallicity dilution and chemical enrichment, but also effects
such as the locking of gas in stars and the launching of galactic winds that work to modulate
the efficiency of these processes.  Instead of searching for
correlations between the changes in the metallicity and these distinct
processes, we vary parameters in our simulations to test their influence
on the evolution of the nuclear metallicity.

%%%%%%%%%%%%%%%%%%%%%%%%%%%%%%%%%%%%%%%%%%%
%%%%%%%%%%%%%%%%%%%%%%%%%%%%%%%%%%%%%%%%%%%
%%%%%%%%%%%%%%%%%%%%%%%%%%%%%%%%%%%%%%%%%%%

\subsubsection{Gas Consumption}
\label{sec:GasConsumption}
Any process that can modify the gas reservoir in the nuclear region
can affect the metallicity measurement.  Here, we consider the
influence of locking gas in the stellar phase when determining the
evolution of the nuclear metallicity.  Star formation lowers
the reservoir of gas in the nuclear region, and hence amplifies the effect of
metallicity dilution that occurs when low metallicity gas floods the
central region.  While this is definitely a physical effect, its
magnitude will be misjudged if the simulations do not properly account
for gas recycling.  Using the stochastic gas recycling method outlined
in \S\ref{sec:GasRecycling}, we are able to modulate the efficiency
with which gas becomes locked in the stellar phase without changing
our star formation efficiency or feedback prescriptions.  Instead, we
change the gas recycling parameter, allowing 
gas to be returned from the stellar state back to the ISM.

We first consider a simulation where newly formed stellar material is 
instantaneously returned to the ISM, such that no gas will be trapped 
in the stellar phase.  At the
opposite extreme, we perform a simulation with star formation, but no
gas recycling (i.e. $\zeta = 0.0$), which would trap the largest
amount of gas in the stellar phase.  In between these extremes, we consider
three values for the gas recycling fraction, $\zeta$, that show how the 
nuclear metallicity evolution changes as the efficiency with which gas is
locked in the stellar state is varied.

\begin{figure*}
  \begin{center}
   \includegraphics[height=2.3in]{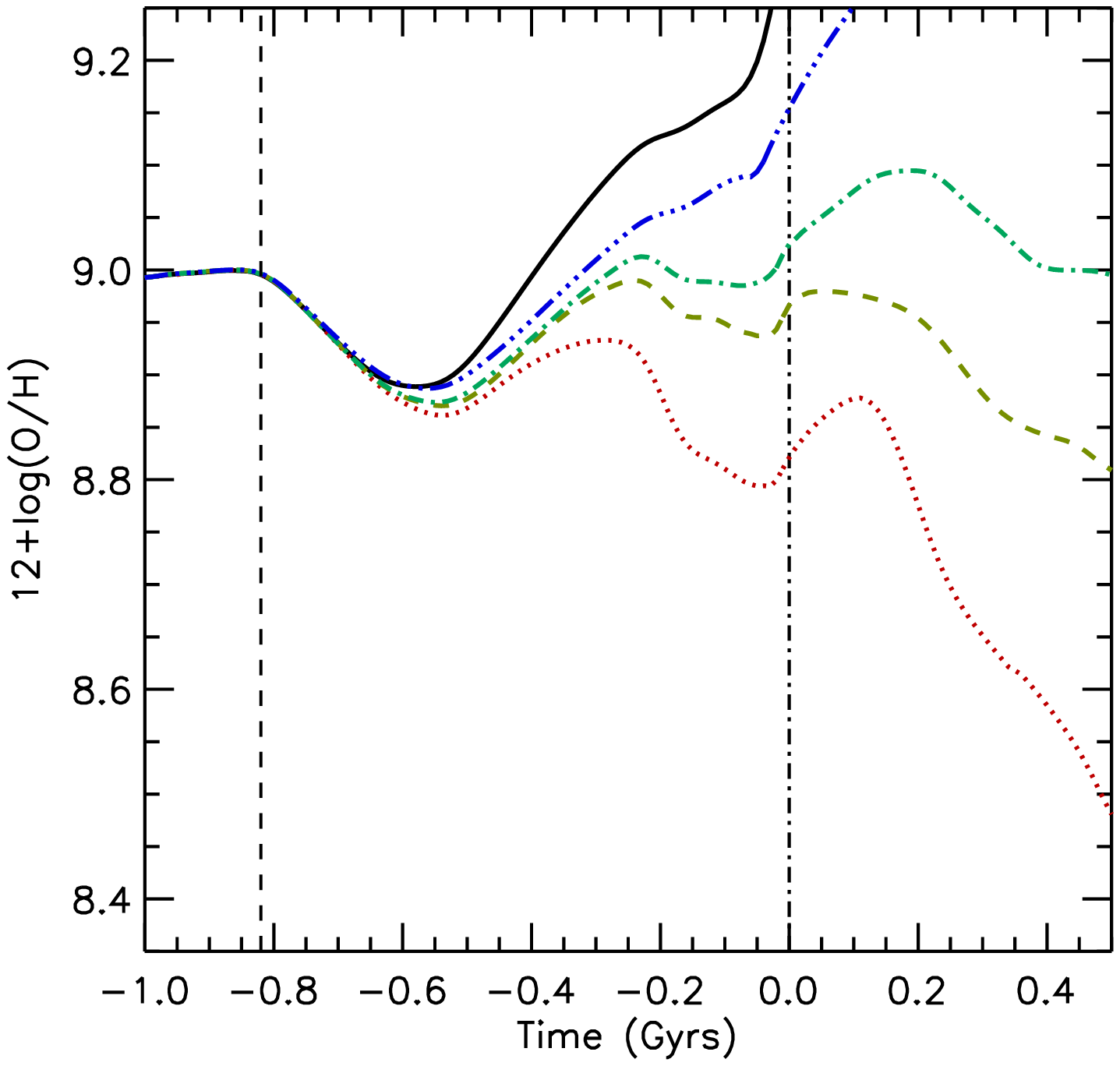}  
   \includegraphics[height=2.3in]{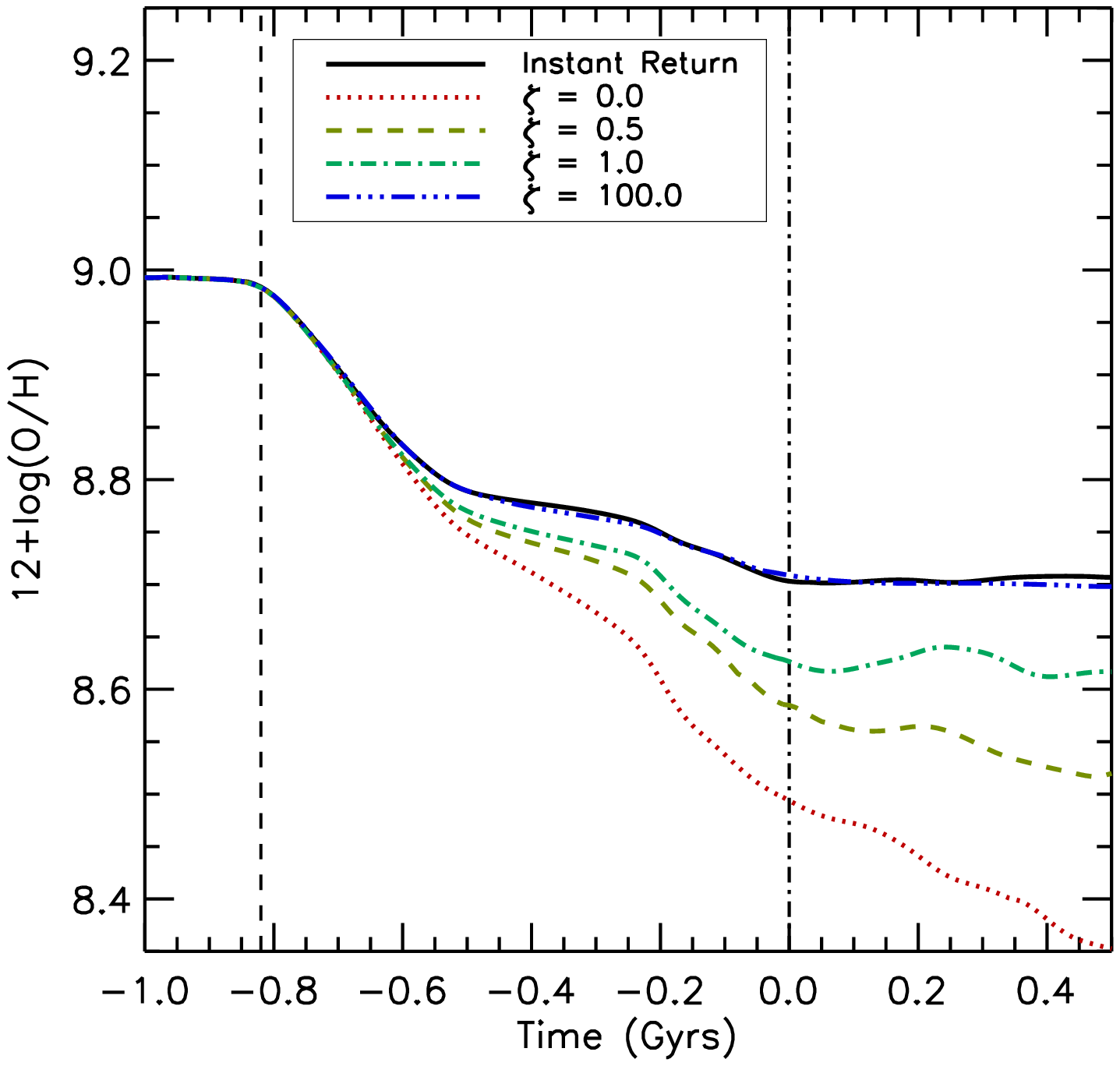}  
   \includegraphics[height=2.3in]{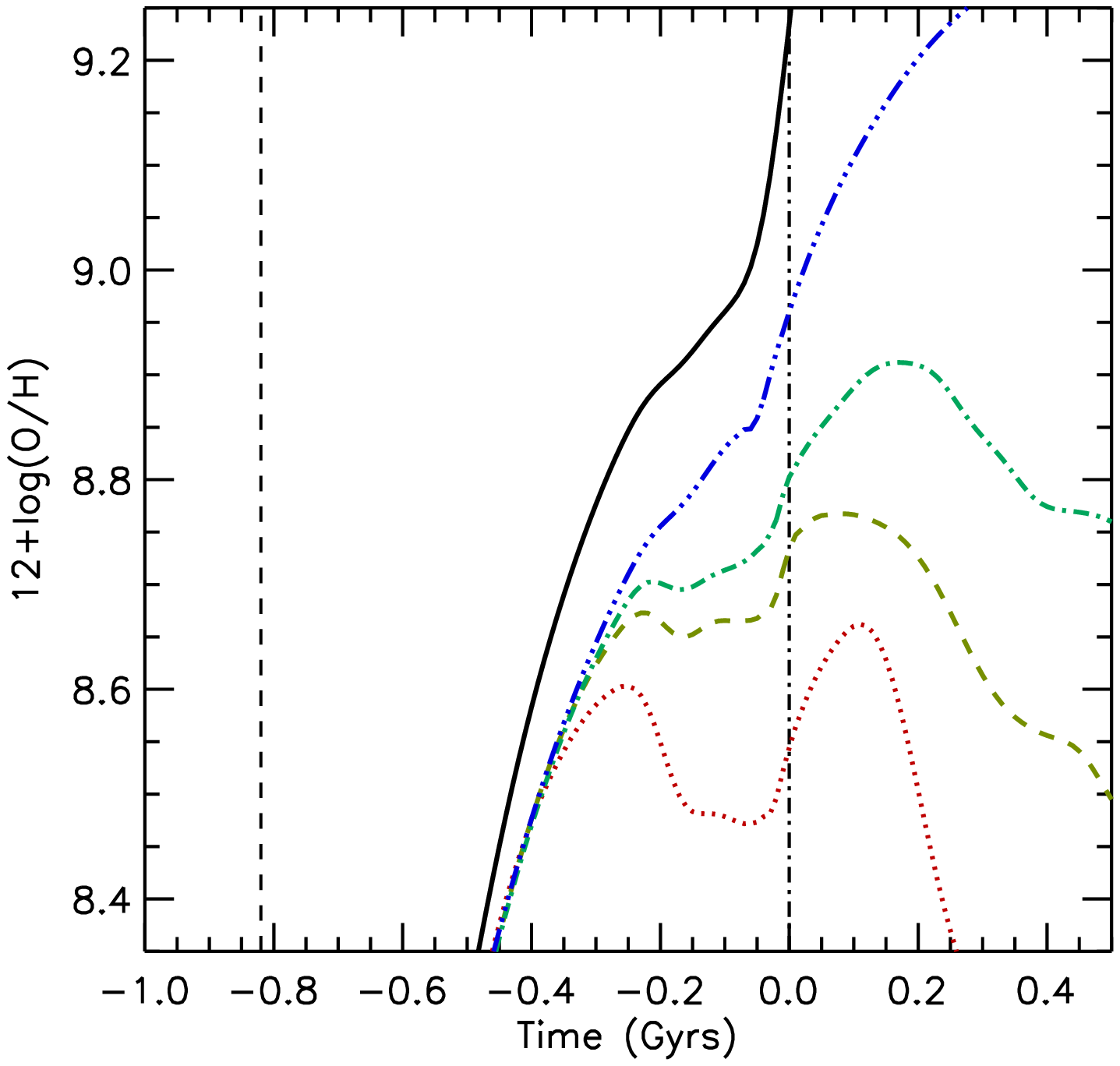}    
\caption{The total (left), dynamical (center), and enriched (right) nuclear metallicities as a function of time for ``BBe'' merger simulations are shown.  The various lines show the nuclear metallicity evolution with several values of the gas recycling coefficient as well as complete and instantaneous gas recycling.  For very large values of the gas recycling coefficient (i.e. $\zeta >> 1$), stellar particles are returned to the gas phase quickly.  As the gas recycling coefficient is increased, the results approach the case of instantaneous and complete gas recycling.  First pericenter passage and final coalescence are marked with vertical dashed and dot-dashed lines, respectively.
}
    \label{fig:ZTpanel}
  \end{center}
\end{figure*}

When the gas recycling parameter is set to very large values, the resulting 
enriched metallicity becomes unphysically high.  This occurs when many 
generations of star formation are allowed to occur over very short timescales
as the gas is quickly recycled into the ISM repeatedly.  However, the resulting 
dynamical metallicity remains physical and meaningful, as it describes the 
change in the nuclear metallicity that will occur as a function of the efficiency with which 
material is locked in the stellar phase.  In fact, dynamical 
metallicity given by our simulation with instantaneous gas recycling (see the black 
line in the center panel of Figure~\ref{fig:ZTpanel}) is very 
similar to the results of~\citet{Rupke2010}.  These simulations capture changes to the nuclear
metallicity caused by dilution without any contributions or contamination from the
creation of stars.  In this case, the nuclear metallicity
is well-correlated with the mass of gas in
the nuclear region, as noted by \citet[][see their Fig. 2]{Rupke2010}.
This indicates that the initially high nuclear metallicity is diluted by
inflows of low metallicity gas.  However, as the gas recycling parameter is dialed 
back to lower values, gas is more efficiently trapped in the stellar phase which increases 
the effect of metallicity dilution.

The simulation with star formation and no gas recycling brackets the
upper end for the amount of gas that can be trapped in the stellar
phase.  In this case, the dynamical metallicity (red dotted line in the center plot of
Figure~\ref{fig:ZTpanel}) is lower than the dynamical metallicities of
all other simulations and falls $\sim$0.2 dex below that of the
nuclear metallicity with instantaneous recycling.  This difference is caused purely
by the effect of locking gas in the stellar state.  When no gas recycling is used, 
a substantial fraction of the inner gaseous reservoir is 
converted into the stellar state, allowing influx of low metallicity gas 
to have stronger influences.  In addition, the enriched
metallicity is lower than any other simulation because the resulting star formation rates are 
lower when gas is efficiently locked in the stellar state.  The resulting total 
metallicity is below the total metallicity where gas recycling is used.

As gas recycling is included at intermediate values, both the dynamical metallicity 
and enriched metallicity increase.  The increase in the dynamical 
metallicity for larger gas return
fractions is well-motivated by the simple physical arguments presented 
above.  Specifically, increasing the gas recycling
fraction lowers the efficiency with gas is depleted from the nuclear 
region into the stellar state.  Hence, as the gas recycling fraction is
increased, the mass of gas in the inner reservoir increases, and the 
influence of metallicity dilution is decreased.

The increase in the enriched metallicity for larger gas return
fractions is similarly explained.  Gas that resides in the nuclear
region will tend to remain there for longer periods of time (on
average) and achieve higher instantaneous and integrated star 
formation rates (hence,
higher metallicities).  While this sheds light on the influence of gas recycling,
we would like to use physically motivated values for the gas recycling
fraction in what follows.  We take the unscaled Starburst99 mass
return fraction as the fiducial mass return fraction used for the remainder of this paper.

\begin{figure*}
  \begin{center}
   \includegraphics[height=2.3in]{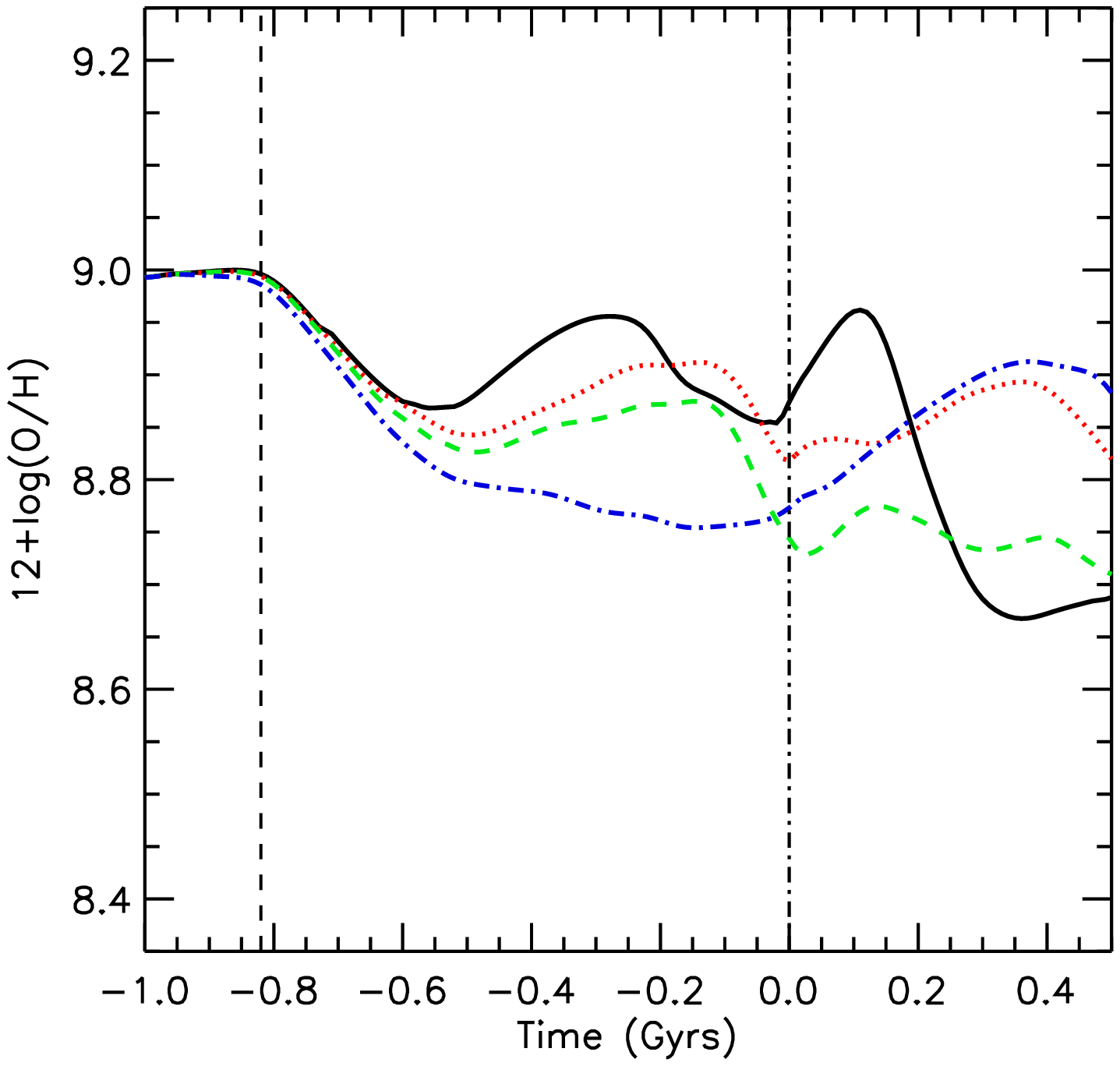}  
   \includegraphics[height=2.3in]{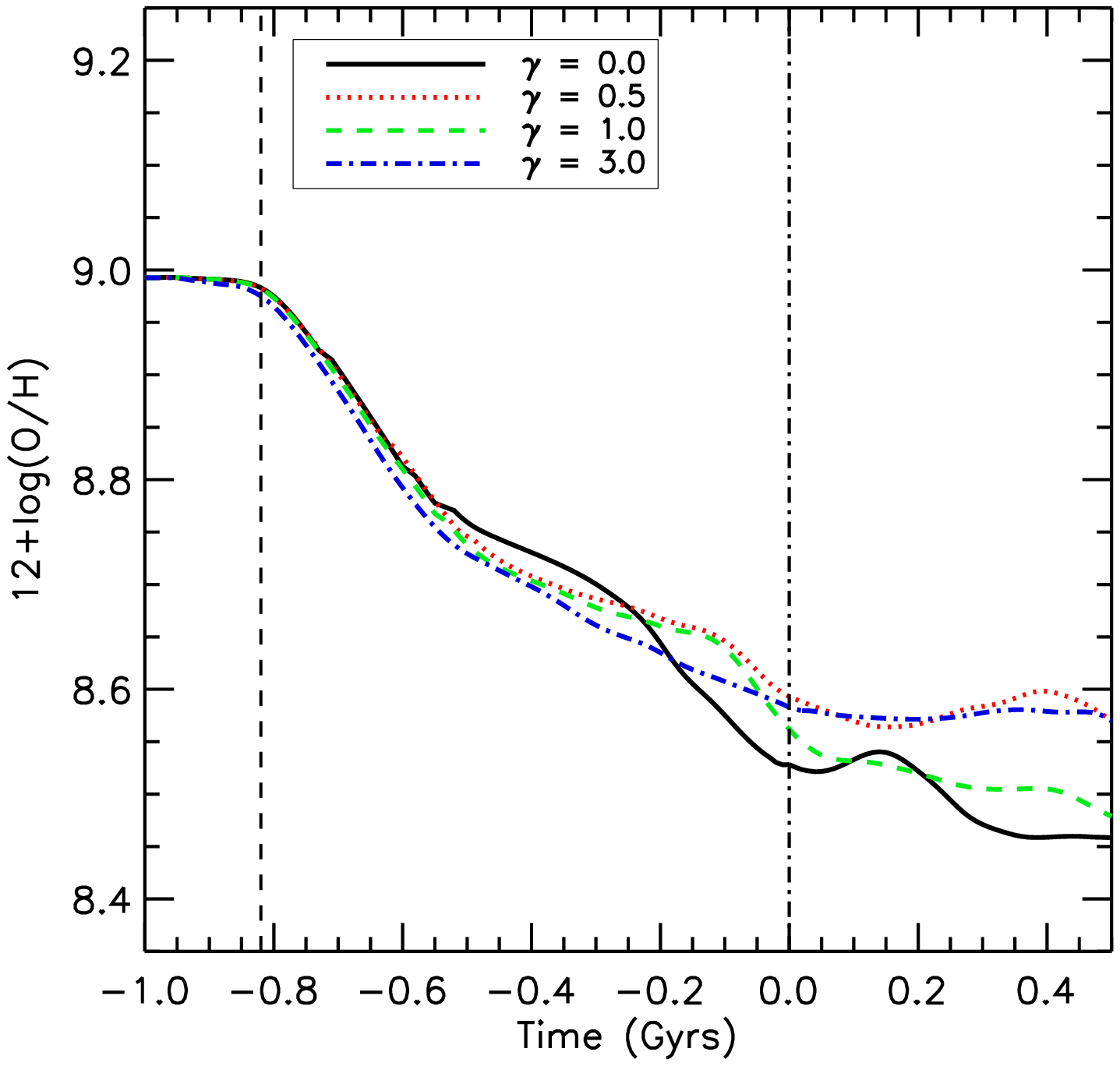}  
   \includegraphics[height=2.3in]{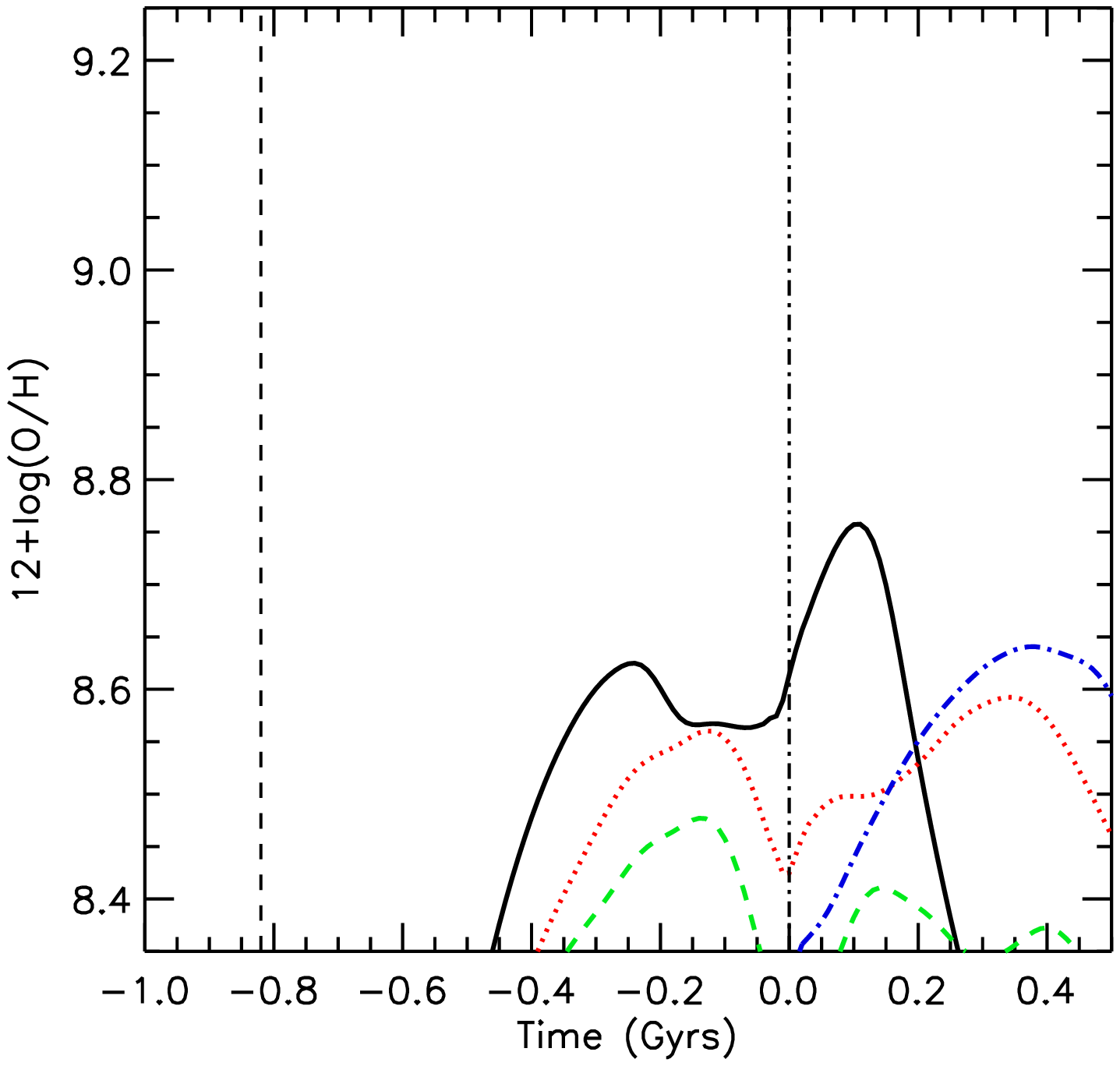}    
   \caption{The nuclear metallicities are shown for varied wind launching efficiencies.    First pericenter passage and final coalescence are marked with vertical dashed and dot-dashed lines, respectively.   }
    \label{fig:WindFraction}
  \end{center}
\end{figure*}

%%%%%%%%%%%%%%%%%%%%%%%%%%%%%%%%%%%%%%%%%%%
%%%%%%%%%%%%%%%%%%%%%%%%%%%%%%%%%%%%%%%%%%%
%%%%%%%%%%%%%%%%%%%%%%%%%%%%%%%%%%%%%%%%%%%

\subsubsection{Galactic Winds}
\label{sec:GalacticWinds}
Galactic winds are thought to play a leading role in the evolution of
the distribution of metals throughout cosmic time.
Starburst--driven, or AGN--driven winds are a ubiquitous phenomena
that act to transport metals from the depths of galaxies into the
interstellar, or intergalactic medium \citep[see, e.g.,][for a
comprehensive review]{VeilleuxARAA}.  Galactic winds may also play a
prominent role in shaping the mass--metallicity relationship
\citep[see, e.g.,][]{DaveMZ2006} and because galaxy mergers induce
episodes of enhanced star formation, they may also induce periods of
significant feedback and galactic outflows thereby altering the
central metallicity evolution of galaxy mergers.  

In the context of this paper, galactic winds can remove gas from central regions and modify the
reservoir of material contributing to the nuclear metallicity~\citep{PerezScoop}.  In this capacity,
the influence of galactic winds is identical to the locking of gas in the 
stellar state.  To demonstrate the impact of galactic winds on the nuclear metallicity, we
present a set of simulations where we vary the mass entrainment,
$\gamma$.  In these simulations we do include gas recycling at the fiducial level, but note
that the our conclusions are unchanged if this effect is turned off.

The dynamical metallicity is not substantially affected by galactic winds.  Since 
star formation is already fairly efficient at locking gas in the stellar state, galactic winds 
provide only an incremental modification to this effect.  However, notable 
changes occur in the enriched metallicity.  Specifically, 
the enriched metallicity is depressed for large values of the mass entrainment because 
gas with high star formation 
rates is either quickly converted into the stellar phase or launched in a wind.  
Therefore, the integrated star formation rates and 
enriched metallicity values for gas particles remains lower, on average, for higher 
mass entrainment values.

The resulting nuclear metallicity, shown in
Figure~\ref{fig:WindFraction}, is depressed as mass
entrainment is increased.  This change is primarily caused by 
changes to the enriched metallicity.  However, there is one additional 
effect that distinguishes galactic winds from locking of material 
in the stellar phase.  While gas that is permanently locked in the stellar state 
will forever be unable to contribute to the nuclear metallicity or 
star formation, gas that has been launched in a wind will ultimately
fall back onto the disk, or even into the nuclear region.  We see this 
effect having an influence in the enriched metallicities of the highest 
mass entrainment value simulations at late times.  Specifically, as gas inflow from 
previously ejected wind material falls upon the nuclear region, the star formation 
rates and enriched metallicities are pushed toward higher values.  

It should be noted that the galactic wind launching prescription used in this work
does not describe a situation where the galactic wind predominately ejects
enriched gas.  In other words, the metallicity of the wind is the same as 
the ambient metallicity of the star forming region from which the wind was launched.  
This prescription could underestimate the efficiency with which metals are pushed 
out of a galaxy and into the intergalactic 
medium.  However, this wind launching prescription is straightforward to understand 
and has the direct and notable effect of removing gas from
the nuclear region. 

Galactic winds do little to change the radial rearrangement of gas
which attends a galaxy merger, as shown by the relative insensitivity
of the dynamical metallicity to the efficiency of the
winds.  The only appreciable effects are on the enriched metallicity.  In 
the following analysis, we take $\gamma = 0.3$
as the fiducial mass entrainment value~\citep{RupkeOutflows2005}.

%%%%%%%%%%%%%%%%%%%%%%%%%%%%%%%%%%%%%%%%%%%
%%%%%%%%%%%%%%%%%%%%%%%%%%%%%%%%%%%%%%%%%%%
%%%%%%%%%%%%%%%%%%%%%%%%%%%%%%%%%%%%%%%%%%%

\subsection{Disk Spin Orientation}
\label{sec:MergerOrientation}

\begin{figure*}
  \begin{center}
   \includegraphics[height=2.3in]{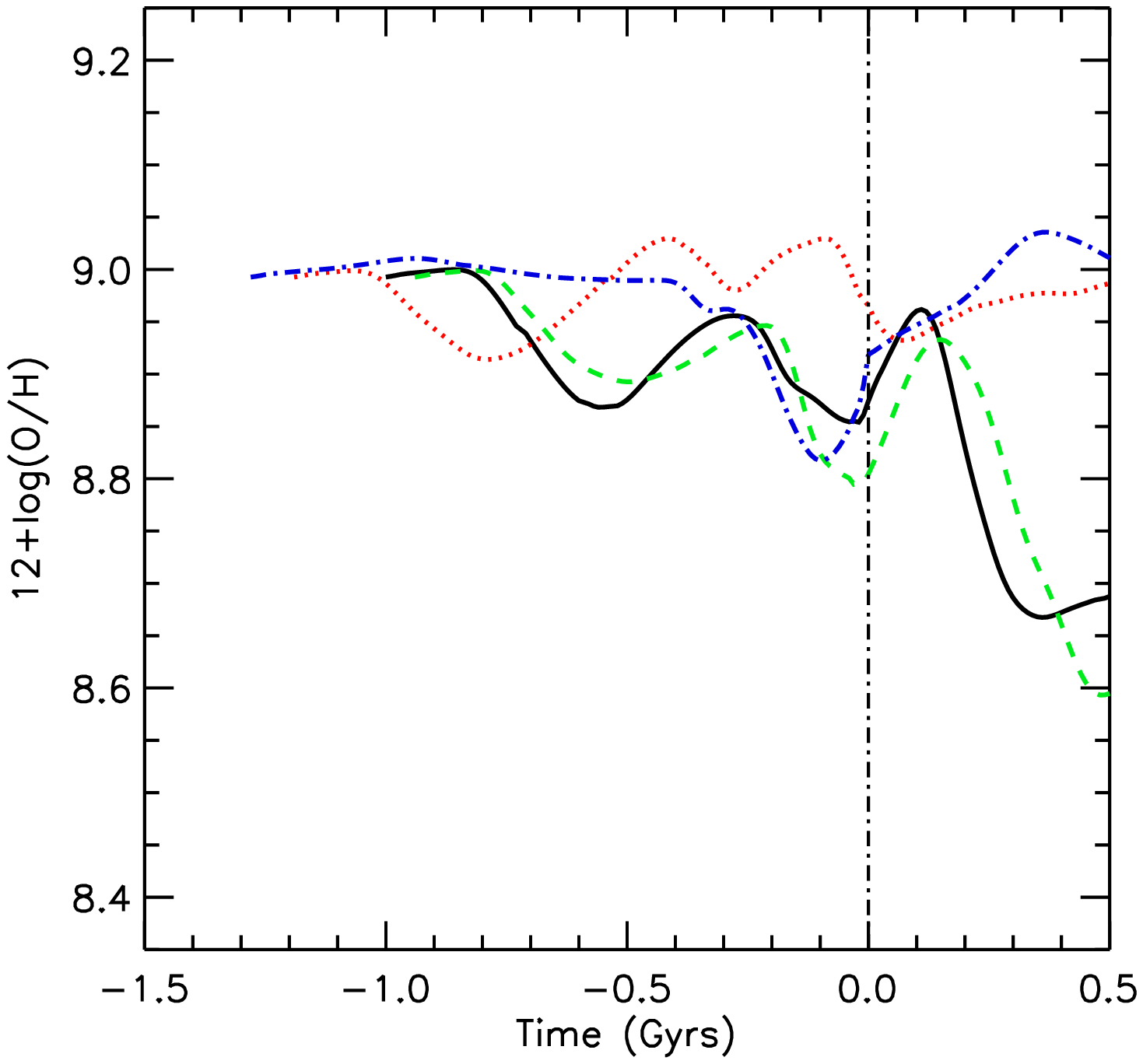}  
   \includegraphics[height=2.3in]{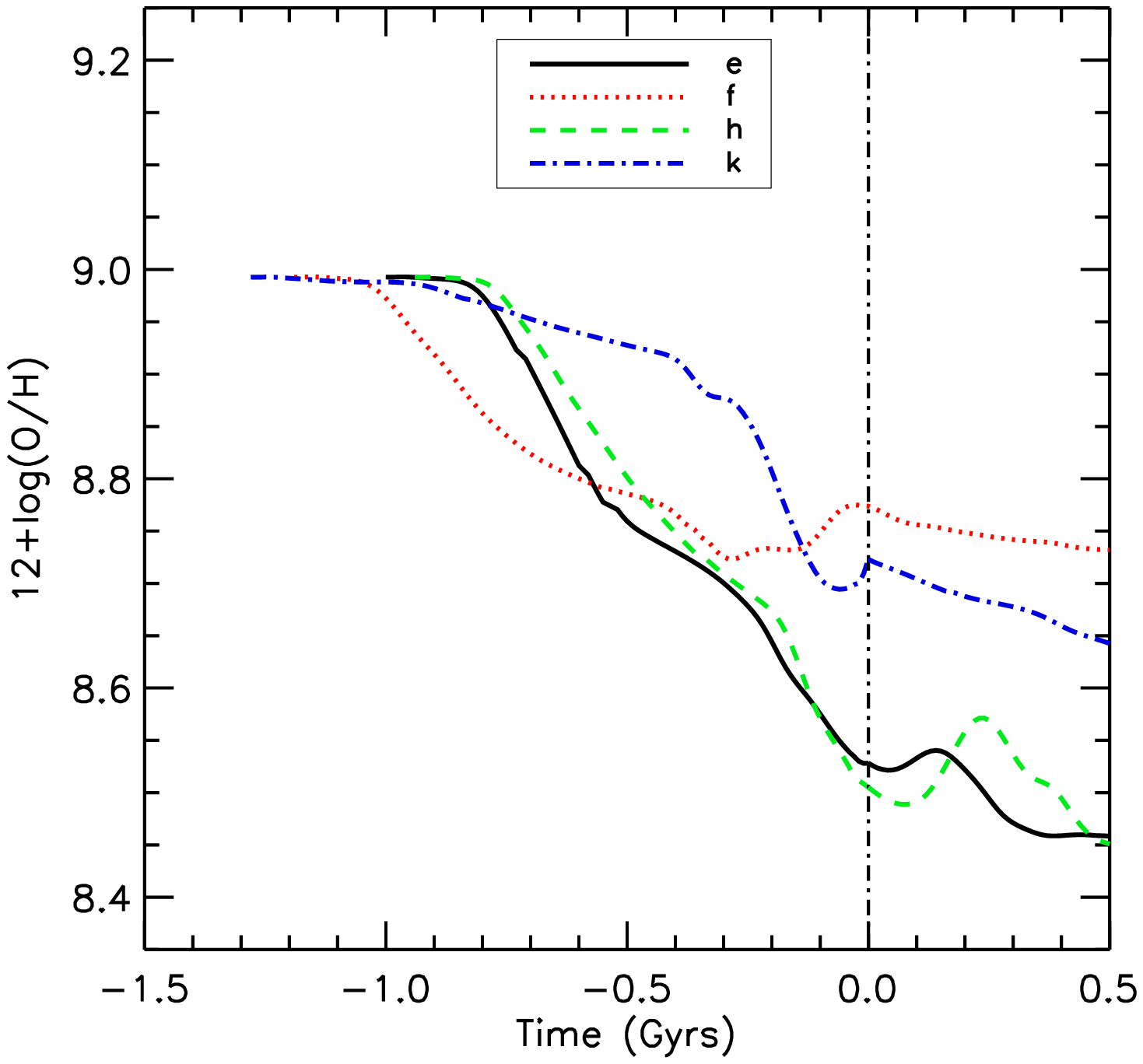}  
   \includegraphics[height=2.3in]{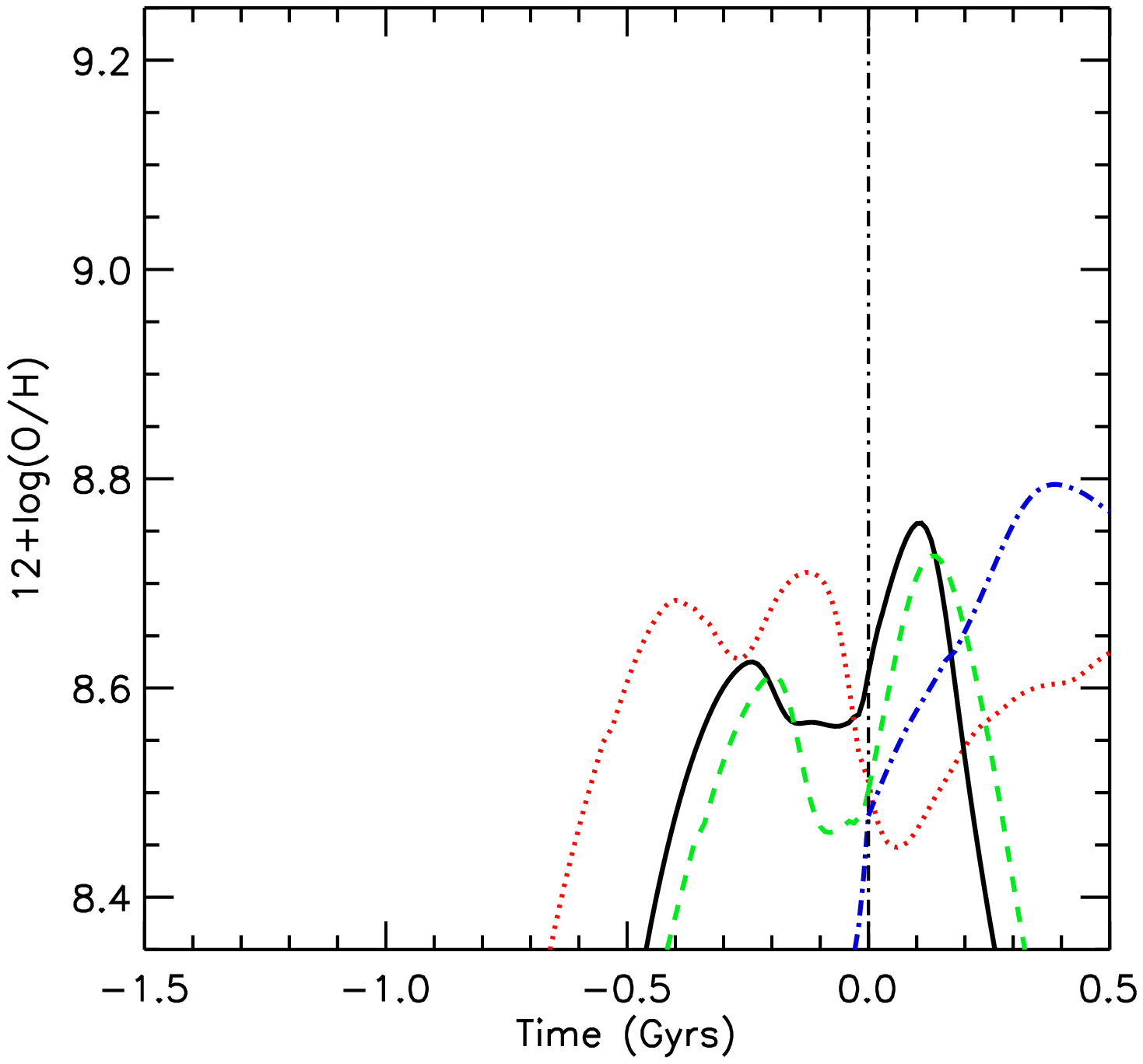}    
   \caption{The nuclear metallicity evolution for four identical initial progenitor disks, set on four 
   distinct merging trajectories, all aligned with final coalescence at $t=0$, as indicated by the dot-dashed vertical line.  Each orientation has a unique metallicity evolution, but all share common features.  First, depressions of the nuclear metallicity are common following any pericentric passage (though not all pericenter passages induce nuclear metallicity depressions, as demonstrated by the ``k'' orientation).  Second, depressions of the nuclear metallicity are common preceding final coalescence.}
   \label{fig:ZTorientations}
  \end{center}
\end{figure*}

While the previous analysis focused on a single merger setup,
the effect of merger orientation can be tested using a set of 
identical ``B'' galaxies merging from 16
orientations (a-p, as detailed in Table~\ref{OrbitalOrientations}).
All orientations maintain unique metallicity evolutions, as the
specifics governing the magnitude and timing of the gas inflows and
starbursts are orientation dependent.  Here we will consider which 
aspects of our fiducial merger's metallicity evolution are found to
hold for various merger orientations.  To test 
this, we first examine the metallicity evolution 
for the four orientations shown in Figure~\ref{fig:ZTorientations}.

\begin{figure}
\begin{center}
   \plotone{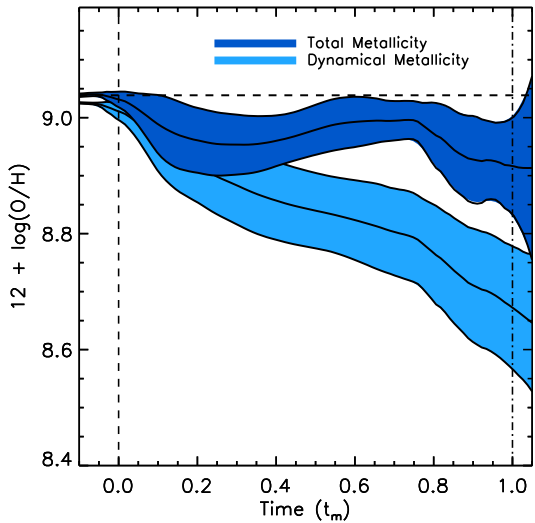}
   \caption{The average metallicity evolution for a set (orientations a-p) of major mergers with identical progenitor galaxies on different merger orientations.  The dark blue region denotes the mean nuclear metallicity plus the one sigma variation for BB galaxy mergers on all 16 orientations.  The light  blue region shows the nuclear metallicity for the same mergers when chemical enrichment from star formation is neglected.  First pericenter passage and final coalescence for all orientations are denoted by the vertical dashed and dot-dashed lines, respectively.    }
    \label{fig:ZTavg}
\end{center}
\end{figure}

As is evident from this figure, three of the four disks show strong
depressions in their nuclear metallicity following pericentric
passage.  This is caused by the strong tidal interaction and resulting
gas inflows which are a natural byproduct of the merger process.  The
disk not experiencing a post pericentric nuclear metallicity
depression is identified as being on a nearly retrograde orientation.
More concretely, the induced gas inflow following pericenter passage
is found to be small compared to the other systems.  The idea that
some orientations, specifically those on retrograde orbits, have
suppressed responses during pericentric passages has been studied
previously \citep{TT72,Donghia10}, and we note that the metallicity
evolution is affected accordingly in the cases where strong
post-pericenter passage responses are not found.

Subsequent depressions in the nuclear metallicity, can be identified
as occurring during periods of strong gas inflow caused by close
encounters.  For the ``e'' and ``h'' orientations, only one additional
close encounter occurs before proceeding to coalescence.  However, for
the ``f'' orientation, three metallicity dips can be seen.  These
metallicity dips are natural products of this particular merger
orientation, which leads to three distinct pericenter passages before
the galaxies ultimately coalesce.  While a hard and fast rule
governing the metallicity depression's dependence on merger
orientation cannot be given, two statements can be made: most
orientations undergo a metallicity depression following pericentric
passages (unless the tidal interaction is sufficiently suppressed), and
most orientations show metallicity depressions immediately preceding
final coalescence.  Both of these statements are demonstrated in figure~\ref{fig:ZTorientations}, 
where three of four (e,f, and h; not k) orientations show post-pericentric dips 
in their nuclear metallicity and three of four (e, k, and h; not f) orientations 
show dips in their nuclear metallicities preceding final coalescence.

If we identify the merging time, $t_m$, to be the time between
pericenter passage and final coalescence, we can average the
metallicity evolution for several orientations to get an orientation
averaged progression of the nuclear metallicity during the merger.
Figure~\ref{fig:ZTavg} shows the mean nuclear metallicity evolution
for all 16 orientations, with the solid region indicating the
1$\sigma$ variations.  Despite the variation that occurs in the
nuclear evolutionary tracks for individual orientations, the mean
result has a well-defined pattern with reasonably small dispersion and
a clear interpretation.  Specifically, there are two notable
metallicity depressions caused by gas inflows, one following
pericentric passage and one preceding black hole coalescence, with a
chemical enrichment induced peak separating in between.  Also shown in
Figure~\ref{fig:ZTavg} is the average dynamical metallicity.  Here we
can see that simulations without chemical enrichment substantially
under-predict the resulting nuclear metallicity.  However, even when
chemical enrichment is included, the mean nuclear metallicity is
depressed throughout the interaction.

%%%%%%%%%%%%%%%%%%%%%%%%%%%%%%%%%%%%%%%%%%%
%%%%%%%%%%%%%%%%%%%%%%%%%%%%%%%%%%%%%%%%%%%
%%%%%%%%%%%%%%%%%%%%%%%%%%%%%%%%%%%%%%%%%%%

\subsection{Progenitor Masses}
\label{sec:MergerMass}

We examine the effect that system mass has on the nuclear metallicity
evolution using a set of major mergers (i.e. 1:1 mass ratios) with
progenitor galaxies of varying mass.  As detailed in \S\ref{sec:GalaxySetup}, 
the four galaxies used in this paper are constructed to be self-similar.  
Hence, the mass study performed here isolates the influence of 
merger mass from other parameters that may scale with mass (e.g. 
gas fraction, galaxy structure, etc.).

\begin{figure}
  \begin{center}
   \plotone{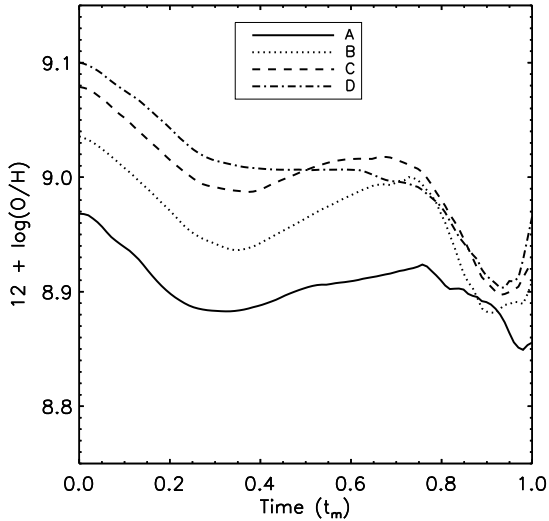}
   \caption{The average metallicity evolution for a set (the e, f, h, and k orientations) of major mergers in four mass bins. }
    \label{fig:ZTmasses}
  \end{center}
\end{figure}

We use four orientations in each
mass bin (the ``e'', ``f'', ``h'', and ``k'' orientations), and
average the resulting metallicity evolutionary tracks.  The result,
shown in Figure~\ref{fig:ZTmasses}, is that the shape of the
evolutionary track is nearly preserved, while the magnitude of the
depression depends weakly on mass.  We note that the preserved shape
of the metallicity evolution supports our argument in the previous
sections that the metallicity evolutionary track is a natural
byproduct of the generic merger process.  This indicates that, within
the limited range of galaxy masses that are used here, the system mass
is not a major consideration in determining the nuclear metallicity
evolution.

%%%%%%%%%%%%%%%%%%%%%%%%%%%%%%%%%%%%%%%%%%%
%%%%%%%%%%%%%%%%%%%%%%%%%%%%%%%%%%%%%%%%%%%
%%%%%%%%%%%%%%%%%%%%%%%%%%%%%%%%%%%%%%%%%%%

\subsection{Gas Fraction}
\label{sec:MergerGasFraction}

\begin{figure*}
  \begin{center}
   \plottwo{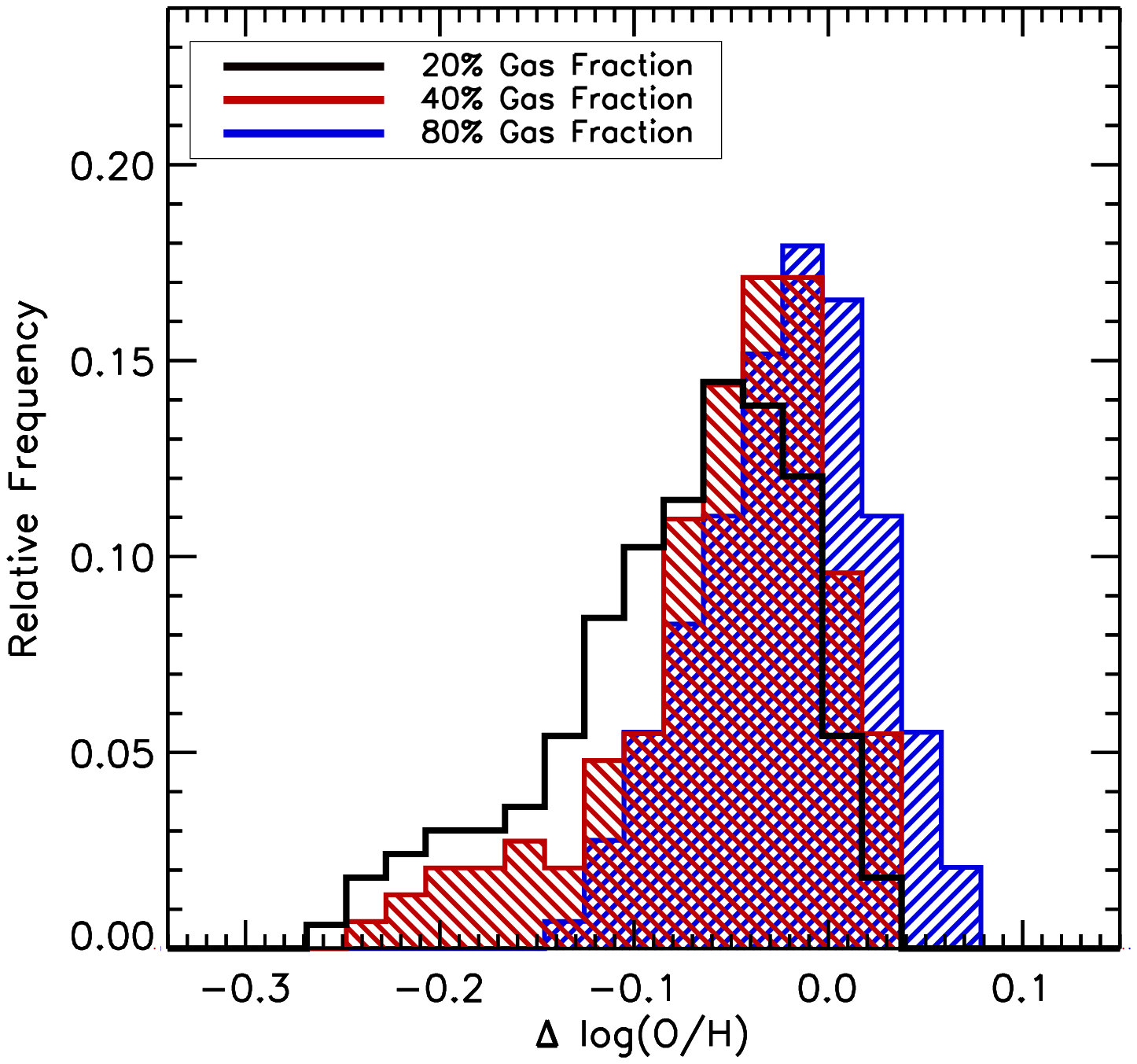}{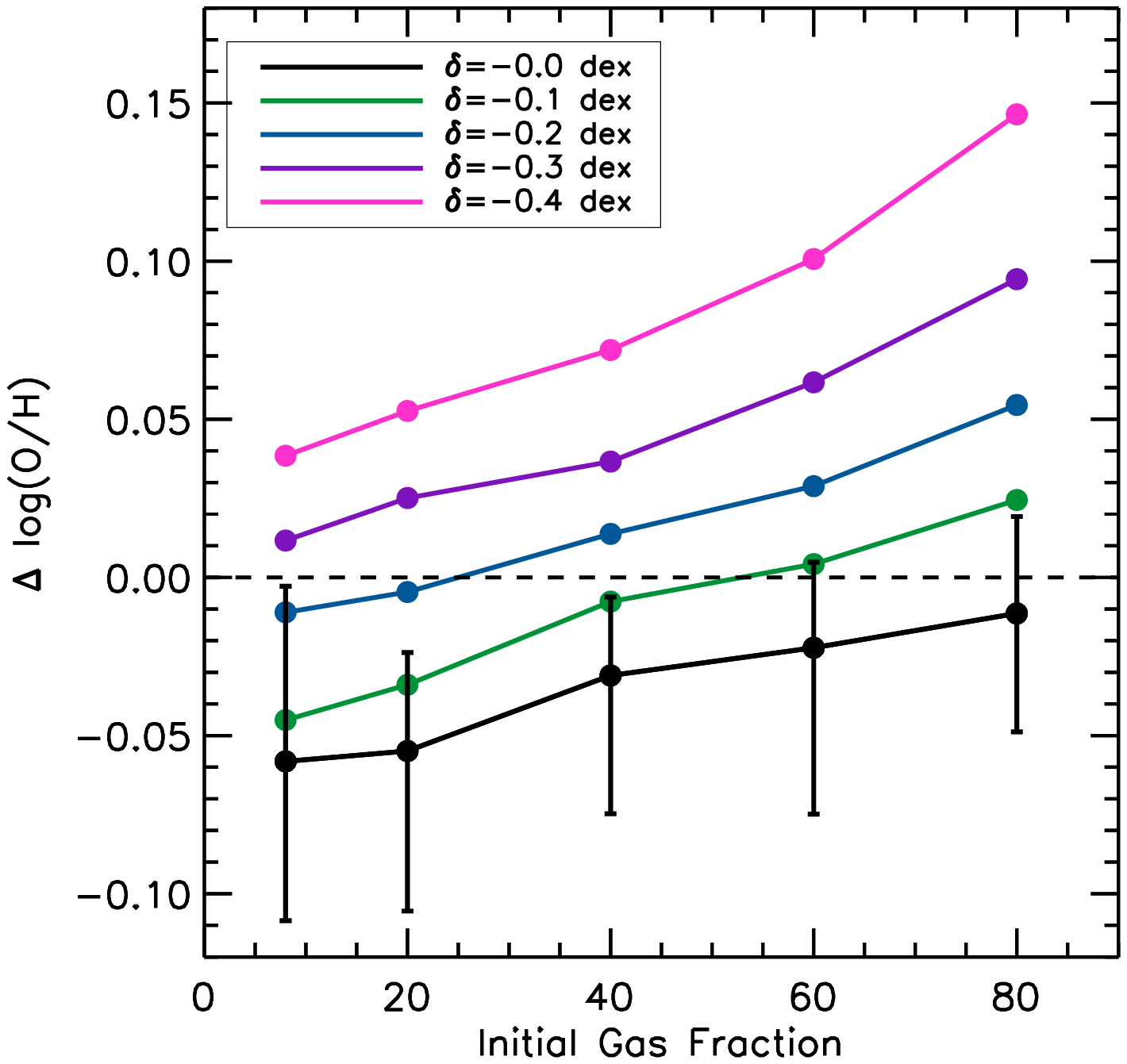}
   \caption{The distribution of changes in the nuclear metallicity for various gas fraction disks (left) and the median change in the nuclear metallicity as a function of gas fraction (right).  The offset to the local MZ relation that was used is given by $\delta$ in the legend (i.e. $\delta=-0.3$ is consistent with the $z > 2$ MZ relation of \citet{ErbMZ2006}). }
    \label{fig:GasFractionHist}
  \end{center}
\end{figure*}

Initial gas fraction is an important parameter to consider in nuclear
metallicity evolution.  For higher gas fraction galaxies that, by definition,
have lower initial stellar mass, two basic changes occur:  
The initial nuclear metallicity of the system, as prescribed by the MZ relation,
is lowered (decreasing the importance of the 
dynamical metallicity), and higher star formation rates occur 
(increasing the importance of chemical enrichment).  The 
combined result, as discussed in this section, 
is that the nuclear metallicities of interacting systems at higher gas fractions 
tend to evolve to larger values, rather than decreasing, 
in contrast to their lower gas fraction 
counterparts.

We first consider the nuclear metallicity evolution for a series of
disks with varied initial gas fractions (8, 20, 40, and 60\%).  We
merge together these systems on the ``e``, ``f'', ``h'', and ``k''
orientations.  If we initialize the metallicity properties of these
disks using the local MZ relation, an assumption that we will relax
later in this section, we can track the metallicity evolution and plot
the change in the nuclear metallicity, as in Figure
\ref{fig:GasFractionHist}.  This shows that the depressions in the
nuclear metallicity that occurred for the low initial gas fractions
considered in the previous sections shift toward less negative, or
even positive values, as the initial gas fraction is raised.  This
transition is caused by the increasing dominance of the enriched
metallicity over the dynamical metallicity, primarily owing to the
higher star formation rates in the higher gas fraction systems.

These highest gas fraction systems, however, are not natural analogs
to galaxies in the local Universe.  Therefore, our choice to
initialize the metallicity properties of these systems using the local
MZ relation comes into question.  Instead, we should consider an MZ
relation better suited for higher redshifts.  A full theoretical
understanding of the origin of the MZ relation has not yet been
obtained, however, cosmological simulations indicate that the shape of
the MZ relation is achieved at early times (i.e. z$>$6) and that
galaxies move along this MZ relation toward higher stellar mass, while
the amplitude of the MZ relation slowly increases over time
\citep{DaveMZ2006}, in broad consistency with observations
\citep{ErbMZ2006,Maiolino2008}.
Without committing ourselves to a single choice for the
high redshift MZ relation, we can re-initialize the metallicity
properties of our galaxies using an MZ relation that is offset by a
constant fixed value, $\delta$, toward lower metallicities.

Since our simulations have no metallicity dependence, re-normalizing the
MZ relation has no effect on the enriched metallicity, but lowers the impact of
the dynamical metallicity.  The result, shown in the right panel of 
Figure~\ref{fig:GasFractionHist}, is that as the the MZ relation is shifted to lower
metallicities, the change in the nuclear metallicity is expected to become
increasingly positive.  For all choices for the initial MZ relation, the higher 
gas fraction systems have a median change in metallicity that is 
more positive than their low gas fraction counterparts.

Using the local MZ relation (i.e. the black line in
Figure~\ref{fig:GasFractionHist}) for systems with relatively low gas
fractions, we find that negative changes to the nuclear metallicity
are expected in a merger.  This is consistent with the observed
lowering of the MZ relation.  However, as we move to higher gas
fractions, or MZ relations better suited for higher redshifts, we find
that increases to the nuclear metallicity are expected.  This result
is caused by shifting the dominant driver of the nuclear metallicity
from metallicity dilution to chemical enrichment.  Moreover, if the
gas content is sufficiently high, the galactic stellar component cannot 
efficiently torque down the gas, resulting in moderated early stage 
inflows \citep{Hopkins09c}.

The same merger
fundamentals apply to low and high gas fraction disks, independent of
their initial metallicities, but the relative increase in the role of
chemical enrichment with respect to the dynamical metallicity causes
higher redshift or gas fraction systems to increase their nuclear
metallicities during the merger process.

Without being overly restrictive about our choice for initial gas
fraction or offset from the local MZ relation, high-redshift, gas-rich
galaxies are expected to have median increases in their nuclear
metallicity when undergoing a merger.  Based on this evidence, we
predict that the depression that occurs in the MZ relation during
galaxy mergers and interactions in the local Universe would not occur
at higher redshift.  Instead, increases in the MZ relation are
expected.

%%%%%%%%%%%%%%%%%%%%%%%%%%%%%%%%%%%%%%%%%%%
%%%%%%%%%%%%%%%%%%%%%%%%%%%%%%%%%%%%%%%%%%%
%%%%%%%%%%%%%%%%%%%%%%%%%%%%%%%%%%%%%%%%%%%

%%%%%%%%%%%%%%%%%%%%%%%%%%%%%%%%%%%%%%%%%%%
%%%%%%%%%%%%%%%%%%%%%%%%%%%%%%%%%%%%%%%%%%%
%%%%%%%%%%%%%%%%%%%%%%%%%%%%%%%%%%%%%%%%%%%

\section{Comparison with Observations}
\label{sec:ComparisonToObservations}

Much work in recent years has gone into observationally classifying
the metallicity of interacting or close pair galaxies in the nearby Universe
~\citep{KGB06,SloanClosePairs,Peeples2009,SolA2010}.  In this section we 
compare our results directly with these observations.  To facilitate this 
comparison, we parameterize the evolution of the nuclear metallicity in 
terms of the galactic projected separation.  We define the projected 
separation as $s = r \cos (\xi)$ with $\xi$ being a uniformly distributed 
random number between $0$ and $2\pi$.  Since all of our systems are, 
by definition, interacting systems, this projected separation neglects
contributions from truly non-interacting systems.  We present the
mass-metallicity (MZ) and separation-metallicity (SZ) relations for a
simulated population of interacting galaxies.

For these comparisons, we generate a population of progenitor galaxies
that contain the metallicity properties of observed field
galaxies~\citep{Shields1990,BelleyRoy1992,Z94,MA04}.  We use a
Gaussian distribution for the nuclear metallicity selection with a
mean value consistent with the MZ relation of~\cite{Tremonti2004} with
a constant standard deviation across mass bins of $\sigma = 0.1$ dex.
The metallicity scale length is taken from a Gaussian distribution
with a mean gradient consistent with~\citet{Z94} of $h_z = h / 0.2$
with $\sigma = 0.3h_z$.

The initial metallicity properties of each galaxy in each snapshot are
assigned based on the above criteria.  Our ability to pick metallicity
properties of the progenitor disks in our post-processing analysis is
a consequence of not having a dynamical dependence on metals or
allowing mixing.  This lets us sample the parameter space of
initial disk metallicity properties thoroughly, without re-running
simulations.

%%%%%%%%%%%%%%%%%%%%%%%%%%%%%%%%%%%%%%%%%%%
%%%%%%%%%%%%%%%%%%%%%%%%%%%%%%%%%%%%%%%%%%%
%%%%%%%%%%%%%%%%%%%%%%%%%%%%%%%%%%%%%%%%%%%

\subsection{The Mass-Metallicity Relation for Interacting Galaxies}

\begin{figure}
  \begin{center}
   \plotone{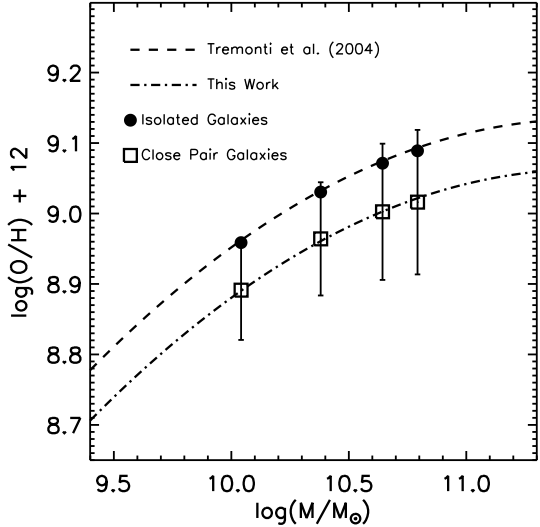}
   \caption{The simulated mass-metallicity relationship for interacting galaxies is shown.  The progenitor galaxy MZ relationship from~\citet{Tremonti2004} is indicated along with the second order best fit from our simulation results. }
    \label{fig:MZ}
  \end{center}
\end{figure}

The MZ relation for our simulated interacting galaxies is shown in
Figure~\ref{fig:MZ}.  We select galaxies with projected separations less than 30
kpc -- the same as was used in~\citet{SloanClosePairs}.
We plot the MZ relation in Figure~\ref{fig:MZ}.  The result is a
depressed MZ relation with an average depression of 0.07 dex, which is
in agreement with the observed depression of 0.05-0.10 dex
\citep{SloanClosePairs}.

Unlike luminosity, which can increase dramatically during 
merger induced starburst activity, the stellar 
mass of an interacting galaxy will change 
only by a modest factor.  Therefore, the evolution 
of galaxies on the MZ relation is driven by changes to the nuclear metallicity, 
not stellar mass.  In the simulation analysis, we are able to quickly determine 
a system's stellar mass.  The stellar masses for the close
pair systems included in the~\citet{SloanClosePairs} sample have 
had their stellar masses determined via optical and IR colors~\citep{KauffmannSDSS,Tremonti2004} 
which have been shown to compare well with the spectrally-determined stellar
masses~\citep{DroryMasses}.  Assuming that the observationally determined stellar 
masses are not strongly affected by ongoing starburst activity, the depression in 
both the observed and simulated MZ relation is being caused entirely by 
changes to the nuclear metallicity -- not by changes in the galactic luminosity or stellar mass.

The depression in our simulated MZ
relation would be slightly smaller if we included a set of field
galaxies that lie on the MZ relation, but appear close in projected
separation (i.e. if we included a population of interlopers).  This 
result gives a comprehensive reproduction of the
depression in the MZ relation for interacting galaxies that accounts
for all relevant physics and the intrinsic scatter in the initial
metallicity properties of the progenitor galaxies, while covering a
reasonable portion of merger phase space.

%%%%%%%%%%%%%%%%%%%%%%%%%%%%%%%%%%%%%%%%%%%
%%%%%%%%%%%%%%%%%%%%%%%%%%%%%%%%%%%%%%%%%%%
%%%%%%%%%%%%%%%%%%%%%%%%%%%%%%%%%%%%%%%%%%%

\subsection{Separation vs Metallicity}
\label{sec:SZ}

\begin{figure}
  \begin{center}
   \plotone{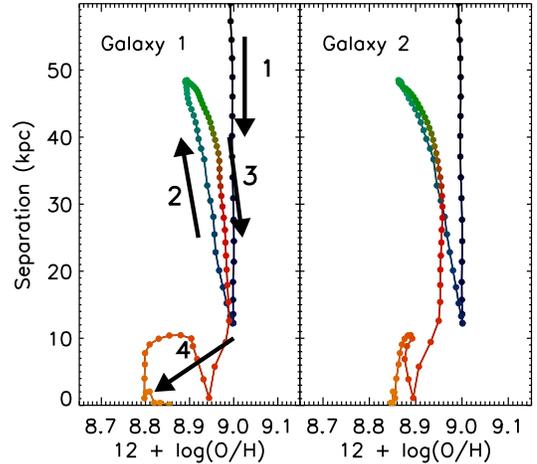}
    \caption{The evolution of two galaxies from one merger simulation are shown.  The numbered arrows follow the evolution of the galaxies as they: 1) Approach first pericenter passage with an unchanging nuclear metallicity, 2) increase their separation while diluting the nuclear metallicity, 3) come together for the second time with 
    notable contributions to the nuclear metallicity from star formation, and 4) lower their nuclear metallicity as they approach final coalescence.  }
    \label{fig:OneSZ}
  \end{center}
\end{figure}

Figure~\ref{fig:OneSZ} shows an example 
of two galaxies (``C'' galaxies) from one merger simulation (the ``e'' 
orientation) moving in separation--metallicity (SZ) space as they 
approach coalescence.  The evolution can be broken down into 
four distinct stages, as marked by arrows in Figure~\ref{fig:OneSZ}.  
First, the galaxies approach first pericenter passage at relatively 
constant nuclear metallicities.  Second, the galaxies separate as gas 
inflows dilute their nuclear metallicity.  Third, the galaxies begin
falling back together, with star formation causing a significant amount 
of chemical enrichment.  And, finally, the galaxies undergo strong 
gas inflows that further dilute their nuclear metallicities as they 
approach coalescence.

The details of this SZ evolution will change 
substantially with the merger orientation.  However, we can gather from 
Figure~\ref{fig:ZTorientations} that all four stages of this 
evolution will be somewhat generic.  Namely, galaxies will present 
constant nuclear metallicities as they approach each other for the first time, and 
strongly depressed nuclear metallicities as they approach final coalescence, with 
mildly depressed nuclear metallicities in between.

Using the evolutionary track of~\ref{fig:OneSZ}, we can develop some expectations 
for the distribution of galaxies in SZ space.  Clearly, we expect close pair galaxies 
to show depressed nuclear metallicities.  In addition, we can expect depressed metallicities 
out to large galactic separations.   While the largest nuclear metallicity 
depressions are expected around close separations, widely separated 
interacting galaxies show mildly depressed nuclear metallicities after 
interacting with their companion.  The observability of these trends can be assessed 
by using our entire merger library and converting real-separations into projected separations.

\begin{figure}
  \begin{center}
   \plotone{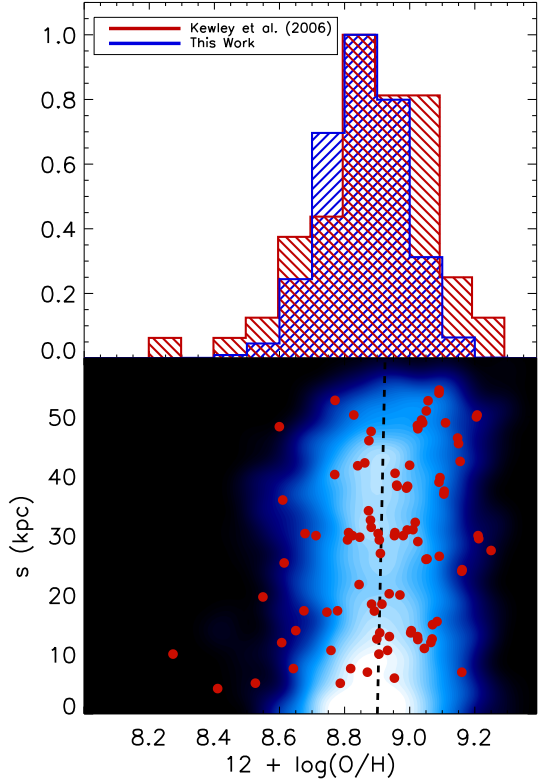}
    \caption{The relative frequency of galaxies in metallicity space (top) and the simulated separation-metallicity relationship along with the data from~\citet{KGB06} (bottom) are shown.  The majority of observed points from~\citet{KGB06} are coincident with our simulated galaxies, with the exception of the low-metallicity, low-separation tail.  Our simulated galaxies show a depression in the nuclear metallicity for close pair galaxies, however this depression is less dramatic than is required to explain the low-metallicity, low-separation tail observed in~\citet{KGB06}.  }
    \label{fig:SZ}
  \end{center}
\end{figure}

The distribution of all simulated galaxies in SZ space
is shown in Figure~\ref{fig:SZ} along with data from~\citet{KGB06}.
The distribution of our simulated galaxies, shown with the colored contours in the 
lower half of Figure~\ref{fig:SZ}, coincide with many of the observed
systems, which is demonstrated in the histogram in the top half of Figure~\ref{fig:SZ}.  
There is a trend toward lower metallicities with close separations.
However, this trend is very mild and the most-extreme low-separation 
low-metallicity objects from the observed sample are not reproduced by our models.  
Despite this difference, it should be noted 
that our simulations agree with the majority of the observed objects
from~\citet{KGB06}, and, there is a systematic but subtle drop in the simulated nuclear
metallicity as the galaxies decrease their projected separation.

We note that the overall distribution of simulated galaxies in SZ space 
is relatively featureless, as evidenced by the contours in Figure~\ref{fig:SZ}.
The nuclear metallicity depressions present at both small and large separations
in Figure~\ref{fig:OneSZ} are washed out as a result of projecting the physical separations,
the wide variety of metallicity dilution versus separations that result from
different spin--orbit couplings, and the scatter in the initial nuclear metallicities as prescribed 
by the mass-metallicity relation.  In addition, while late stage mergers (e.g., stage 4 in 
Figure~\ref{fig:OneSZ}) present the largest depressions in nuclear metallicity, 
they only last for a short period of time compared to the overall merger process.  Since 
we place equal weight on every timestep for every simulation when determining the 
distribution of simulated galaxies in Figure~\ref{fig:SZ}, the highly-depressed late-stage 
mergers will naturally have a less dramatic impact.

Several factors can influence the distribution of our model galaxies
in SZ space.  In particular, the (in)completeness of our merger
library and the metal enrichment scheme will influence the resulting
SZ relation.  We discuss both of these factors in the following
subsections, so that we may better understand the disagreement between
our models and the observations at low separations and low
metallicities.  However, we also note that additional observations
would help clarify how common these low-separation, low-metallicity
objects are.

\subsubsection{Merger Library Completeness}

Although the merger library used in this work was fairly comprehensive, there is a 
significant portion of merger parameter space that has not been 
explored.  Perhaps this could explain why we do not reproduce the
low-separation, low-metallicity objects.
We can describe our exploration of merger parameter 
space with two components:  Variations of initial isolated galaxies models and 
merger configuration parameters.

Additional merger configurations (i.e. variations to the orbital orientation and 
angular momentum) are unlikely to yield surprising results.  Out of all of our 
merger simulations, a maximum depression of about 0.3 dex was found, with 
many depressions being must less drastic.  To achieve the lowest-metallicity 
objects found in~\citet{KGB06}, depressions of $\sim$ 0.6-0.7 dex would be required.  
It seems unlikely that the isolated galaxy models used in this study will reproduce very low
metallicity objects, even with a more extensive exploration of merger orientations.

On the other hand, galaxy models that are significantly different from 
the ones used in this study, such as low 
mass galaxies, could fill out a new region in SZ space.  For example, 
galaxies with initial stellar masses of $M_{s} \sim 10^9 M_{\odot}$ will have 
initial nuclear metallicities of $12+ \log (O/H) \sim 8.6$ and could have their 
nuclear metallicities depressed to $\sim$ 8.3 or 8.4 during the merger.
More generally, it seems likely that the lowest metallicity objects of~\citet{KGB06}
are likely merging systems that had low metallicities prior to the onset of the merger which may be
a reasonable possibility given the low B-band luminosities associated with these objects.  
These systems may be reproduced by our simulations 
when we include a more extensive 
set of initial galaxy models.

\subsubsection{Approximating Delayed Mixing}

Although oxygen abundances are
well-approximated by an instantaneous enrichment model, this does not
necessarily imply that metals produced in SNe instantaneously migrate
to HII regions.  Specifically, we can relax our assumption that the
metallicity of an SPH particle should be instantaneously identical to
the metallicity of an HII region.  Our analysis calculated the nuclear
metallicity from the current metallicities of gas particles.  While
the metals produced in core collapse supernovae are quickly returned
to the ISM ($\sim 10^7$ years) and supernovae remnants cool and fade
into the ambient ISM over similar timescales ($\sim 10^6$ years),
there could then be some additional delay before this gas is
sufficiently mixed with the surrounding medium and ends up in HII
regions.  In particular, we expect the metals produced by SNe to
affect the \textit{next} generation of HII regions.  As a first
approximation, this does not require a change to our simulations, but
rather a change to our post-processing analysis.

\begin{figure}
  \begin{center}
   \plotone{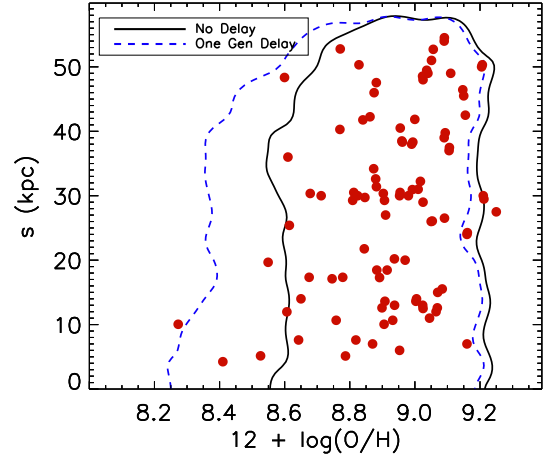}
    \caption{The simulated separation-metallicity relationship is shown using a delayed chemical enrichment scheme.  Newly formed metals are allowed to contribute to the HII region metallicity measurements only after the SPH particle in which they reside undergoes an integrated star formation rate equal to its own mass.  This effectively delays metal enrichment by one generation. }
    \label{fig:MultiContourSFR}
  \end{center}
\end{figure}

\begin{figure}
  \begin{center}
   \plotone{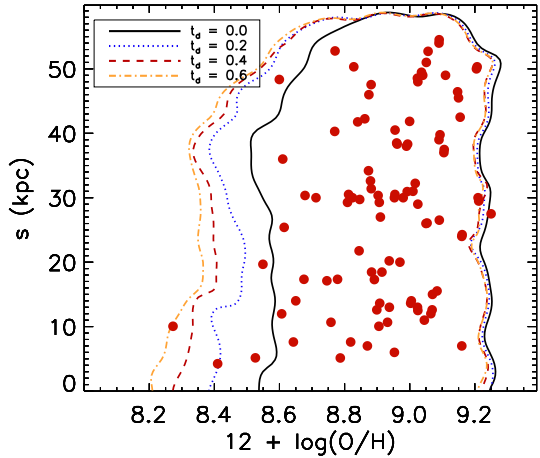}
    \caption{The simulated separation-metallicity relationship is shown using a delayed chemical enrichment scheme.  Newly formed metals are allowed to contribute to the HII region metallicity measurements after a specified delay time, $t_d$, in Gyrs.   }
    \label{fig:MultiContour}
  \end{center}
\end{figure}

We can define an effective HII region metallicity that is given by 
the metallicity that the SPH particle had one generation of star 
formation ago.  If the number of generations of star formation is given by
\begin{equation} 
N_* (t)  = \frac{1}{M} \int _0 ^t \frac{dM_*}{ dt^\prime} dt ^\prime
\end{equation}
where $M$ is the particle mass, then the effective HII region metal content is given by
\begin{equation}
 M_{Z} ^\prime (t) = \left\{
  \begin{array}{l l}
    M_{Z,init} + y M \left( N_* - 1 \right) & \quad \text{if $N_* \ge 1$ }\\
    & \\
    M_{Z,init} & \quad \text{if $N_* < 1$ }\\
  \end{array} \right.
\end{equation}
which reflects the particle's metallicity one generation of star formation in the past.  

This delayed chemical enrichment scheme gives an SZ relation, shown in
Figure~\ref{fig:MultiContourSFR}, which has a low metallicity tail
that is qualitatively similar to the observed galaxies.  This does not
imply that our previous calculation of the nuclear metallicity using
the non-delayed chemical enrichment values were incorrect, but rather
that the assumption of instantaneous transition of metals into HII
regions may be overly restrictive.  This is a plausible explanation,
provided that the associated delay-times are reasonable estimates for
the timescale over which SN ejecta would mix and diffuse into the ISM.

If we parameterize an effective HII region metallicity in terms of a delay-time,
rather than an integrated star formation rate, we can get an estimate
for the required delay time.  Rather than neglecting the enrichment from the last
generation of star formation, we neglect the enrichment
that occurred over the past $t_d$ years.  The effective metal content of a
particle is then given by
\begin{equation}
 M_{Z} ^\prime (t) = \left\{
  \begin{array}{l l}
    M_{Z,init} + y \int _0 ^ {t - t_d} \frac{dM_*}{dt^\prime} dt ^\prime & \quad \text{if $t \ge t_d$ }\\
    & \\
    M_{Z,init} & \quad \text{if $t < t_d$ }\\
  \end{array} \right.
\end{equation}
where $t_d$ is the assumed time delay.  

This time delayed chemical enrichment scheme gives an SZ relation,
shown in Figure~\ref{fig:MultiContour}, where larger delay times lead
to systematic decreases in the inferred metallicity of close pairs.
This is similar to those seen in Figure~\ref{fig:MultiContourSFR}, but
with an explicit time dependence.  The delay times required for our
simulations to match observations is of the order of a few $10^8$
years, which is comparable to estimates for the metal mixing timescale
of newly produced metals into the ISM~\citep{OeyMetalMixing,
ScaloElmegreenMetalMixing}.  We conclude that delaying the transfer of
metals from SN remnants to HII regions is important to consider;
however, it is unclear whether these long mixing times are appropriate
for the extreme environments in merging galaxy nuclei where external forces, 
gas inflows, high gas densities, and large star formation rates are
present.

%%%%%%%%%%%%%%%%%%%%%%%%%%%%%%%%%%%%%%%%%%%
%%%%%%%%%%%%%%%%%%%%%%%%%%%%%%%%%%%%%%%%%%%
%%%%%%%%%%%%%%%%%%%%%%%%%%%%%%%%%%%%%%%%%%%

\section{Discussion}
\label{sec:Discussion}

\subsection{Predictions for Galaxies with High Gas Content}

The results of \S~\ref{sec:MergerGasFraction} demonstrate that the
merger--induced metallicity evolution is heavily dependent upon
gas content, specifically, higher gas fraction disks produce less
nuclear metallicity depression than lower gas fraction disks (see, e.g., Fig.~\ref{fig:GasFractionHist}).
In the extreme, high gas content mergers may even yield significant
{\it enhancements} of central metallicity, a result that emerges both because
of the increased impact of enrichment from star formation
as well as because nuclear gas
inflows are suppressed when the gas fraction is very high
\citep{Hopkins09c}.

As a consequence of these trends, we expect that the nuclear
metallicities of gas--rich merging galaxies will be similar to,
or greater than, an appropriately matched quiescent sample.  This is in
contrast to the depression of nuclear metallicity observed in low
gas--content mergers which was the focus of this paper.   Such considerations
may be applicable at high redshift where current observations suggest that
galaxies contain significantly higher gas fractions \citep{Tacconi2010,ErbGasFraction}
and suppressed metallicity at fixed galaxy mass \citep{ErbMZ2006,Maiolino2008} or for
less--massive gas--rich galaxies in the local universe \citep{Cat10}.  

The predicted metallicity enhancements in gas--rich merging systems was also
noted in a recent paper by \citet{PerezScoop} using similar methods.   While
both works observed similar metallicity enhancements, the timing was not identical,
and purported cause of these effects were very different.  The  \citet{PerezScoop}
model shows rapid metallicity enhancement during early stages of the merger primarily
as a result of rapid star formation owing to disk instabilities in the  gas--rich system
(see, e.g., their Figs.~8 and 9).  Such effects occur prior to, or during the early
stages of the merger and are therefore decoupled from the final coalescence stage.  In
our model, the gas--rich disks are designed to be stable a priori, and therefore the
metallicity enhancements are driven by intense star formation triggered by the merger.
While the end result is the same, namely metallicity enhancements, the relative timing
and strength of the vigorous star formation, the metallicity dilution owing to inflows,
and the metallicity enhancement owing to star formation are fundamentally different, and
therefore potentially discernible via observations.

We caution, however, that the galaxy merger simulations carried out here 
may require some generalization before they are directly applicable 
at high  redshifts.  Effects such as the higher UV background, higher merger rate, and direct  fueling of low 
metallicity gas via the surrounding intergalactic medium at high redshifts may
be of significant influence in the observed metallicity of galaxies.   And numerical
uncertainties, such as the proper treatment of feedback from star formation and
accreting black holes will invariable need to be explored.

\subsection{Future Considerations: An Improved Chemical Enrichment Model}
\label{sec:future}

By comparing our models of interacting galaxies to the observed mass-metallicity (MZ)
and separation--metallicity (SZ) relations we have refined our
understanding of the factors at play during the metallicity evolution of galaxies.
It is crucial to point out (yet rarely done so), however, that numerical results such
as ours are based upon the ability to robustly track a number of complicated and interrelated
astrophysical processes.  While we consider our sub--resolution models to be well
motivated and well tested, we admit that there are straightforward improvements to
these models which will allow for more stringent tests of our physical picture.  We
now take a moment to outline these future improvements to our models.

The chemical enrichment model employed here tracks
a global metallicity through a single scalar variable that is continuously updated
assuming that the entire solar yield (assumed to be a mass fraction of 0.02) is
instantaneously recycled within a single star--forming fluid element (i.e.,
an SPH particle).  We admittedly note the simplicity of this model, but
consider the ease with which it is implemented, the straightforward analysis
it allows, and its inherent self--consistency as distinct advantages of
this approach.  The instantaneous recycling approximation is also justified
because we compare to oxygen derived metallicities, and oxygen is thought
to be recycled on a short ($\sim 10^7$ years) timescale through core--collapse
supernovae.  Moreover, in a practical sense, similar models have been employed
in nearly every SPH code over the last two decades which makes our results easy
to gauge within the existing theoretical framework.

There are, however, avenues in which distinct progress can be
made to our chemical enrichment model.  Efforts toward this end
have been pioneered by a number of authors over the last $\sim$5-10 years
\citep[see, e.g.,][to name a few]{KawataGasReturn,TornatoreGasReturn,
ScannapiecoGasReturn,OppenheimerandDave,KimAbel} by employing two
common features: metallicity--dependent cooling, and more complex
enrichment schemes that explicitly track various species and their
time--dependent release.  Our gas recycling scheme is
particularly well-suited for including mass return by a single
stellar population.  As individual star particles are converted back
to interstellar gas, they can straightforwardly carry with them the
appropriate enrichment from Type I and II supernovae as well as
mass--loss from massive young stars and asymptotic giant branch stars.
Such an approach is necessary to study trends in abundance ratios,
such as those observed in the centers of elliptical galaxies.

Additional accuracy may also be gained by including sub--resolution models for
the turbulent diffusion of metals \citep[see, e.g.,][]{WadsleyDiffusion,ShenDiffusion}.  
Coupled to an advanced enrichment algorithm, a physically motivated procedure for  
diffusing metals will allow enrichment to occur locally but still have global influences
if conditions warrant.  
Such methods have been shown to be important for tracking metallicity in cosmological
simulations \citep[e.g.,][]{ShenDiffusion} and we intend to assess their
influence in isolated and merging galactic systems.

Finally, the comparisons between the models and the observations can be significantly
improved by actually tracking the detailed line emission using 3D radiative transfer
\citep{JonssonCoxGroves}.  As efforts to produce data cubes with full spatial and
spectral information are developed~\citep[e.g.][]{RichKewley2010}, full metallicity 
maps can be produced and directly compared to our simulations.  Radiative 
transfer in our simulations will shed light on any additional information that may
be encoded in the line profiles and allow for a more even--handed comparison.

\section{Conclusion}

Using numerical simulations, we have investigated the impact that mergers
have on galactic nuclear metallicity.  Our models include the capability of
describing star formation, chemical enrichment, gas recycling, and star--
formation driven winds.  The analysis performed here relies upon the ability
to accurately track the spatial and temporal return of metals to the ISM
enabled by our stochastic gas--recycling algorithm within GADGET (see
\S~\ref{sec:GasRecycling}). 

One of the primary results of this work is to reinforce the notion that
galaxy mergers, and their attending gravitational tidal forces, generate
significant inflows of gas shown explicitly in Figures~\ref{fig:ZTorientations}
and \ref{fig:ZTavg} ~\citep[see also][]{Rupke2010, DilutionPeak2010, PerezScoop}.
These inflows transport metal--poor gas into the nuclear region and reinforce
the generic connection between close galaxy interactions and nuclear metallicity
dilution -- no matter how many close passages occur.

Merger--induced gaseous inflows are, however, not the only factor influencing the
nuclear metallicity, and this paper quantifies the competing effects of
enrichment from star formation (\S~\ref{sec:DilutionAndEnrichment}), gas
consumption (\S~\ref{sec:GasConsumption}), and galactic winds
(\S~\ref{sec:GalacticWinds}).  Our models provide a physical template to
understand nuclear metallicity evolution and lead us to conclude that
merger--induced metallicity dilution and chemical enrichment from star formation
drive the nuclear metallicity, while gas consumption and galactic winds play a
secondary role to modulate the efficiency of the primary processes.

We have made a distinct effort in this work to directly compare our models to
observations, allowing us to validate the physical evolutionary model that we
have presented and associate observable trends with the various stages of the
merger evolution.   We find that the merger--induce depressions of nuclear
metallicity are typically $\sim0.07$ dex, an offset similar to that observed
to the mass--metallicity relation for close pairs in SDSS~\citep{SloanClosePairs}.
The models also demonstrate that central depressions of metallicity are time--dependent 
owing to the specific merger orbit, the galaxies' spin--orbit coupling,
and varying structural properties of the galaxies.  When compared to the observed
separation--metallicity relation of~\citet{KGB06}, the models generally show a 
similar trend to produce lower central metallicities at smaller separations.  The
precise role of interactions is difficult to discern, however, because of the
significant scatter in the relations (see \ref{fig:SZ} and the discussion in~
\S\ref{sec:SZ}).  Future comparisons in which galaxies are morphologically
classified according to merger stage, or separated according to IR luminosity or
star--formation rate may be better suited to isolated the true effects of the
interaction.

This work also demonstrates that the progenitor gas content can have
a profound influence on the resulting merger--induced metallicity dilution.
Mergers between gas--rich progenitors can yield systematic metallicity
enhancement, as opposed to metallicity dilution (see \S~\ref{sec:MergerGasFraction},
and specifically Fig.~\ref{fig:GasFractionHist}).  In our picture,  
central metallicity enrichment comes from vigorous merger--induced star formation that can
compensate for, and eventually overcome the merger--induced inflow of  
metal--poor gas.  This view is slightly different than put forth in recent  
work by \citet{PerezScoop}, which argues that disk instabilities and clump formation drive rapid  
star formation and subsequent metal enrichment.  But once these processes occur,  
which is typically early in the merger process, the story becomes similar to when the  
galaxies have low gas content.  By studying the detailed correlations between merger stage,  
pair separation, central metallicity, and metallicity gradients, in both observations  
and the models, future work will allow for a better understanding of the (merger and  
enrichment) processes at work.

%%%%%%%%%%%%%%%%%%%%%%%%%%%%%%%%%%%%%%%%%%%
%%%%%%%%%%%%%%%%%%%%%%%%%%%%%%%%%%%%%%%%%%%
%%%%%%%%%%%%%%%%%%%%%%%%%%%%%%%%%%%%%%%%%%%

\bibliography{NMpaper}{}
\bibliographystyle{apj}

\end{document}